\documentclass[letterpaper,11pt]{article}
\pdfoutput=1

\usepackage{jheppub}
\usepackage{multirow}
\usepackage{shuffle}

\usepackage{subfig}
\usepackage{xspace}
\usepackage[countmax]{subfloat}
\usepackage{enumitem}
\usepackage{amssymb}
\usepackage{amsmath}
\usepackage{cancel}
\usepackage{color}
\usepackage{braket}
\usepackage{graphicx}
\usepackage{multirow}
\usepackage{verbatim}
\usepackage{amsthm}
\usepackage{slashed}
\usepackage{wasysym}
\usepackage{simplewick}
\usepackage{mathtools}
\usepackage{soul}
\usepackage{xspace}

\newcommand{\fd}[2]{\parbox{#1}{\includegraphics[width=#1]{#2}}}

\newcommand{\setcaptionskip}{\setlength\baselineskip{14pt}}
\newcommand{\setmainskip}{\setlength\baselineskip{18pt}}

\def\cA{\mathcal{A}}
\def\cB{\mathcal{B}}
\def\cC{\mathcal{C}}

\def\cL{\mathcal{L}}
\def\cM{\mathcal{M}}
\def\cN{\mathcal{N}}
\def\cO{\mathcal{O}}

\def\cP{\mathcal{P}}

\def\nn{{\nonumber}}

\newcommand{\hard}{\mathrm{hard}}
\newcommand{\dyn}{\mathrm{dyn}}

\newcommand{\bare}{\mathrm{bare}}

\newcommand{\Eq}[1]{Equation~\eqref{#1}}

\DeclareRobustCommand{\Sec}[1]{Sec.~\ref{#1}}

\DeclareRobustCommand{\Fig}[1]{Fig.~\ref{#1}}

\DeclareRobustCommand{\Eq}[1]{Eq.~(\ref{#1})}

\def\be{\begin{equation}}
\def\ee{\end{equation}}

\newcommand{\nbar}{{\bar n}}

\newcommand{\Sl}[1]{\slashed{#1}}

\allowdisplaybreaks[3]

\newcommand{\sdt}{\!\cdot\!}

\newcommand\bn{{\bar n}}

\newcommand{\la}{\lambda}

\newcommand{\dbar}{d\hspace*{-0.08em}\bar{}\hspace*{0.1em}}

\preprint{\vbox{\hbox{MIT-CTP 4933}\hbox{CALT-TH-2017-055}}}

\title{Fermionic Glauber Operators and Quark Reggeization}

\author[1,2]{Ian Moult,}
\author[3]{Mikhail P. Solon,}
\author[4]{Iain W. Stewart,}
\author[4]{and Gherardo Vita}

\affiliation[1]{Berkeley Center for Theoretical Physics, University of California, Berkeley, CA 94720, USA}
\affiliation[2]{Theoretical Physics Group, Lawrence Berkeley National Laboratory, Berkeley, CA 94720, USA}
\affiliation[3]{Walter Burke Institute for Theoretical Physics, California Institute of Technology, Pasadena, CA 91125, USA}
\affiliation[4]{Center for Theoretical Physics, Massachusetts Institute of Technology, Cambridge, MA 02139, USA}

\emailAdd{ianmoult@lbl.gov}
\emailAdd{mpsolon@caltech.edu}
\emailAdd{iains@mit.edu}
\emailAdd{vita@mit.edu}

\abstract{We derive, in the framework of soft-collinear effective field theory (SCET), a Lagrangian describing the $t$-channel exchange of Glauber quarks in the Regge limit. The Glauber quarks are not dynamical, but are incorporated through non-local fermionic potential operators. These operators are power suppressed in $|t|/s$ relative to those describing Glauber gluon exchange, but give the first non-vanishing contributions in the Regge limit to processes such as $q\bar q \to gg$ and $q\bar q \to \gamma \gamma$. They therefore represent an interesting subset of power corrections to study. 
The structure of the operators, which describe certain soft and collinear emissions to all orders through Wilson lines, is derived from the symmetries of the effective theory combined with constraints from power and mass dimension counting, as well as through explicit matching calculations. Lightcone singularities in the fermionic potentials are regulated using a rapidity regulator, whose corresponding renormalization group evolution gives rise to the Reggeization of the quark at the amplitude level and the BFKL equation at the cross section level. We verify this at one-loop, deriving the Regge trajectory of the quark in the $3$ color channel, as well as the leading logarithmic BFKL equation. Results in the $\bar 6$ and $15$ color channels are obtained by the simultaneous exchange of a Glauber quark and a Glauber gluon. SCET with quark and gluon Glauber operators therefore provides a framework to systematically study the structure of QCD amplitudes in the Regge limit, and derive constraints on higher order amplitudes.}

\begin{document} 

\maketitle

\section{Introduction} \label{sec:intro}

The study of limits of amplitudes and cross sections plays an important role in our understanding of gauge theories by providing constraints on higher order calculations, as well as a glimpse at the all orders structure of the theory. One limit that has been intensely studied since the early days of field theory, both in QED~\cite{Gell-Mann:1964aya,Mandelstam:1965zz,McCoy:1976ff,Grisaru:1974cf} and QCD~\cite{Fadin:1975cb,Kuraev:1976ge,Lipatov:1976zz,Kuraev:1977fs,Balitsky:1978ic,Lipatov:1985uk,Lipatov:1995pn}, is the Regge or forward limit, $|t| \ll s$. The simplicity of this limit lead to the discovery of integrability in QCD~\cite{Lipatov:1993yb,Faddeev:1994zg}, and allows for an understanding at finite coupling in $\cN=4$ super Yang-Mills theory~\cite{Bartels:2014mka,Basso:2014pla,Sprenger:2016jtx}. In this limit large logarithms, $\log(s/|t|)$, appear in the perturbative expansion at weak coupling, and their resummation dresses the $t$-channel propagator, leading to an amplitude that behaves as $(s/|t|)^{\omega}$, where $\omega$ is the Regge trajectory. This behavior is referred to as Reggeization, and directly predicts terms in the higher order perturbative expansion of amplitudes, placing important constraints on their structure (see e.g.~\cite{Dixon:2011pw,Dixon:2014iba,Dixon:2015iva,Caron-Huot:2016owq,Dixon:2016nkn} for applications). The Regge trajectory for the gluon is known to two loops in QCD~\cite{Fadin:1995xg,Fadin:1996tb,Korchemskaya:1996je,Blumlein:1998ib,Fadin:1998py}, and to three loops in non-planar $\cN=4$ \cite{Henn:2016jdu}. Recently there has been progress in understanding the breaking of naive Reggeization, and Regge-cut contributions, leading to a more complete picture of forward scattering at higher loops~\cite{Bret:2011xm,DelDuca:2011ae,Caron-Huot:2013fea,Caron-Huot:2016tzz,Caron-Huot:2017fxr}. At the cross section level the resummation is described by the Balitsky--Fadin--Kuraev--Lipatov (BFKL) equation~\cite{Kuraev:1977fs,Balitsky:1978ic}. 

A powerful approach for studying the limits of gauge theories is the use of effective field theory (EFT) techniques. The framework of soft collinear effective theory (SCET)~\cite{Bauer:2000ew, Bauer:2000yr, Bauer:2001ct, Bauer:2001yt} has been widely used to study the soft and collinear limits of QCD, including power suppressed contributions in these limits (see e.g.~\cite{Larkoski:2014bxa,Moult:2016fqy,Kolodrubetz:2016uim,Moult:2017rpl,Feige:2017zci}). Recently an EFT for forward scattering~\cite{Rothstein:2016bsq} was developed in the framework of SCET, providing a systematic way of analyzing the Regge limit at higher perturbative orders and at higher powers in the expansion in $|t|/s$. In~\cite{Rothstein:2016bsq}, the leading power operators that describe the exchange of $t$-channel Glauber gluons were derived, and it was shown that their rapidity renormalization~\cite{Chiu:2012ir,Chiu:2011qc} gives rise to amplitude level Reggeization and the cross section level BFKL equation. For other approaches to studying the subleading power corrections in the Regge limit see~\cite{Amati:1987wq,Amati:1987uf,Amati:1990xe,Amati:1992zb,Amati:1993tb,Akhoury:2013yua,Luna:2016idw}.

In this paper we apply the EFT for forward scattering to the Reggeization of the quark. This is interesting for a number of reasons.
First, quark exchange in the $t$-channel provides the leading contribution for certain flavor configurations in $2\to 2$ forward scattering in QCD, such as $q \bar{q} \to gg$ and $q \bar{q} \to \gamma \gamma$, and is thus important for understanding the behavior of such amplitudes.
Second, the Reggeization of the quark is power suppressed relative to that of the gluon, and therefore provides a simple case for studying the structure of SCET at subleading power in the Regge limit. Third, the application to quark Reggeization further develops the operator based framework, which together with~\cite{Rothstein:2016bsq} provides a description of the Regge limit for both quark and gluon exchanges which seamlessly interfaces with the standard SCET for the study of hard scattering.

The study of the Reggeization of the quark has a long history. In QED, the photon does not Reggeize due to the abelian nature of the theory, but the electron does, providing the first field theoretic derivation of Regge phenomenon~\cite{Gell-Mann:1964aya,Mandelstam:1965zz,McCoy:1976ff,Grisaru:1974cf}. The BFKL equation for $e^+e^-\to \gamma \gamma$ has also been studied in QED~\cite{Sen:1982xv}. In QCD, the Reggeization of the quark has received less attention since it is at subleading power compared to the Reggeization of the gluon. It was first studied in~\cite{Fadin:1976nw,Fadin:1977jr}, and Reggeization was proven to leading logarithmic (LL) order in~\cite{Bogdan:2006af}. Under the assumption of Reggeization, the two-loop Regge trajectory for the quark was derived in~\cite{Bogdan:2002sr} from the next-to-next-to-leading order $2\to2$ scattering amplitudes in QCD~\cite{Anastasiou:2000kg,Anastasiou:2000ue,Anastasiou:2001sv,Glover:2001af,Bern:2002tk}. Interestingly, to this order it is the same as the Regge trajectory of the gluon, up to so-called Casimir scaling, i.e. replacing $C_A \to C_F$.

The emphasis of this paper is the development of the EFT framework for forward scattering, with the hope of facilitating progress in understanding the structure of the Regge limit of QCD. 
We derive the operators describing the $t$-channel exchange of a Glauber quark in the Regge limit. These operators are fixed by the symmetries of the effective theory, constraints from power and mass dimension counting, and explicit matching calculations. They describe certain soft and collinear gluon radiation to all orders, and have not previously appeared in the literature. For a single emission, they reduce to the vertex of Fadin and Sherman~\cite{Fadin:1976nw,Fadin:1977jr}, which is the analogue of the Lipatov vertex~\cite{Kuraev:1976ge} for the case of a Reggeized quark. 
As a demonstration of our framework, we verify explicitly at one-loop that the rapidity renormalization of our potential operators leads to the Reggeization of the quark at the amplitude level and to the BFKL equation at the cross section level, thus providing another LL proof of these results but in the modern language of renormalization. We also show that it is simple to derive results for amplitudes in the $\bar 6$ and $15$ color channels by considering the simultaneous exchange of a Glauber quark and a Glauber gluon.

An outline of this paper is as follows. In \Sec{sec:review} we briefly review the formulation of SCET with Glauber gluon operators from~\cite{Rothstein:2016bsq}.  In \Sec{sec:ferm_glaub} we derive the structure of the fermionic Glauber operators. We consider Glauber quark exchanges between two collinear particles as well as between a collinear and a soft particle, and discuss their power counting. We also give the relevant Feynman rules.
In \Sec{sec:tree_level} we perform a tree level matching calculation onto the operators, which is sufficient to fix their precise form to all orders in $\alpha_s$. In \Sec{sec:quark_reggeize} we derive the one-loop Reggeization of the quark using the rapidity renormalization of the operators. We also show that rapidity finite contributions arising from box graphs with both a Glauber quark and a Glauber gluon reproduce known results in the $\bar 6$ and $15$ channel. In \Sec{sec:quark_BFKL} we derive the BFKL equation for $q \bar{q} \to \gamma \gamma$, and show that it is equivalent to the standard BFKL equation up to Casimir scaling. We conclude and discuss future directions in \Sec{sec:conc}.

\section{SCET with Glauber Operators}\label{sec:review}
In this section we briefly review the structure of SCET with Glauber operators, following~\cite{Rothstein:2016bsq}. This also allows us to define the notation used throughout the paper. We will gloss over many subtleties in the construction of the effective theory, and refer the interested reader to~\cite{Rothstein:2016bsq} for a more detailed discussion.   

SCET is an effective theory of QCD that describes the interactions of collinear and soft particles~\cite{Bauer:2000ew, Bauer:2000yr, Bauer:2001ct, Bauer:2001yt, Bauer:2002nz}. Let us focus on the single lightlike direction relevant for 2 to 2 forward scattering (multiple lightlike directions are considered in~\cite{Rothstein:2016bsq}). We define two reference vectors $n^\mu$ and $\bn^\mu$ 
such that $n^2 = \bn^2 = 0$ and $n\sdt\bn = 2$. Any momentum $p$ can then be written as
\begin{equation} \label{eq:lightcone_dec}
p^\mu = \bn\sdt p\,\frac{n^\mu}{2} + n\sdt p\,\frac{\bn^\mu}{2} + p^\mu_{\perp}\
\,.\end{equation}
A particle is referred to as ``$n$-collinear'' if it has momentum $p$ close to the $\vec{n}$ direction, or more precisely, if the components of its momentum scale as $(n\!\cdot\! p, \bn \!\cdot\! p, p_{\perp}) \sim 
(\la^2,1,\la)$. Here $\la \ll 1$ is a formal power counting parameter, which is determined by the scales defining the measurement or kinematic limits. We will write the SCET fields for $n$-collinear quarks and gluons, as $\xi_{n}(x)$ and $A_{n}(x)$. In addition to describing collinear particles, SCET also describes soft particles, which have momenta that scale as $(\lambda,\lambda, \lambda)$, and are described in the EFT by separate quark and gluon fields, $q_{s}(x)$ and $A_{s}(x)$. This theory is sometimes called SCET$_\text{II}$~\cite{Bauer:2002aj}.

The SCET Lagrangian is expanded as
\begin{align} \label{eq:SCETLagExpand}
\cL_{\text{SCET}}=\cL_\hard+\cL_\dyn= \cL^{(0)} + \cL_G^{(0)} +\sum_{i\geq0} \cL_\hard^{(i)}+\sum_{i\geq1} \cL^{(i)}\,,
\end{align}
with each term having a definite power counting, ${\cal O}(\lambda^i)$, denoted by the superscript. As written, the SCET Lagrangian is divided into three different contributions. The $ \cL_\hard^{(i)}$ contain hard scattering operators, and are derived by a matching calculation, and are process dependent. The $\cL^{(i)}$ describe the long wavelength dynamics of soft and collinear modes in the effective theory, and are universal. The leading power Glauber Lagrangian $\cL_G^{(0)}$ describes interactions between soft and collinear modes in the form of potentials, which break factorization unless they can be shown to cancel. It is derived in~\cite{Rothstein:2016bsq} and discussed below.

Operators in SCET are formed from gauge invariant building blocks. The gauge invariant $n$-collinear quark and gluon fields are defined as
\begin{align} \label{eq:chiB}
\chi_{{n}}(x) &= \Bigl[W_{n}^\dagger(x)\, \xi_{n}(x) \Bigr]
\,,\qquad 
\cB_{{n}\perp}^\mu(x)
= \frac{1}{g}\Bigl[ W_{n}^\dagger(x)\,i  D_{\perp}^\mu W_{n}(x)\Bigr]
 \,,
\end{align}
with analogous definitions for $\bn$-collinear fields.
The collinear Wilson line is given by
\begin{align}\label{eq:Wilsonline}
W_n =\left[  \sum\limits_{\text{perms}} \exp \left(  -\frac{g}{\bar \cP } \bar n \cdot A_n(x)  \right) \right]\,,
\end{align}
where $\cP$ is the so-called label operator, which picks out the large component of a given momentum. These operators involve non-local Wilson lines, but are local at the scale of the dynamics of the EFT. The gauge invariant soft fields are defined in a similar manner, with
\begin{align}\label{eq:gauge_soft}
\cB_{S\perp}^{\bar n \mu}=\frac{1}{g}[S_\bn^\dagger i D^\mu_{S\perp} S_{\bar n}]\,, \qquad \cB^{n\mu}_{S\perp}=\frac{1}{g} [S_n^\dagger i D^\mu_{S\perp} S_n]\,.
\end{align}
These operators involve Wilson lines of soft gluons, and are non-local at the soft scale.

The leading power Glauber Lagrangian in SCET$_\text{II}$~\cite{Rothstein:2016bsq} is given by
\begin{align}\label{eq:Glauber_Lagrangian}
\cL_G^{\text{II}(0)} 
&= e^{-ix\cdot \cP} \sum\limits_{n,\bar n} \sum\limits_{i,j=q,g}   \cO_n^{iB} \frac{1}{\cP_\perp^2} \cO_s^{BC}   \frac{1}{\cP_\perp^2} \cO_{\bar n}^{jC}   + e^{-ix\cdot \cP} \sum\limits_{n} \sum\limits_{i,j=q,g} \cO_n^{iB}   \frac{1}{\cP_\perp^2} \cO_s^{j_n B}\,,
\end{align}
which gives contributions that scale as $\cO(\lambda^0)$. Glauber modes are not dynamical in the EFT but are incorporated through $\frac{1}{\cP_\perp^2}$ potentials, which are instantaneous in the light cone directions and non-local in the $\perp$ direction. 
In Eq.~(\ref{eq:Glauber_Lagrangian}) the first term describes the scattering of $n$ and $\bar n$ collinear particles, while the second term describes the scattering of collinear particles with soft particles. This Lagrangian is exact and does not receive matching corrections in $\alpha_s$ since no hard interactions are being integrated out~\cite{Rothstein:2016bsq}. Moreover, iterated potentials are reproduced by time ordered products ($T$-products) in the effective theory.

Each term in Eq.~(\ref{eq:Glauber_Lagrangian}) is written in a factorized form with gauge invariant operators that sit at different rapidities. The $n$-collinear operators are given by
\begin{align}
\cO_n^{qB} = \bar \chi_n T^B \frac{\Sl{\bar n}}{2} \chi_n \,, \qquad \cO_n^{gB} = \frac{i}{2} f^{BCD} \cB^C_{n\perp \mu} \frac{\bar n}{2} \sdt (\cP+\cP^\dagger ) \cB^{D\mu}_{n\perp}\,,
\end{align}
with $\bar n$-collinear operators identical under the replacement $n\leftrightarrow \bar n$. The soft operators are given by
\begin{align}\label{eq:iain_ira_soft}
\cO_s^{BC}&=8\pi \alpha_s \bigg \{   \cP^\mu_\perp S_n^\dagger S_{\bar n} \cP_{\perp \mu} -\cP^\perp_\mu g\tilde \cB_{S\perp}^{n\mu} S_n^\dagger S_{\bar n} - S_n^\dagger S_{\bar n} g \tilde \cB_{S\perp}^{\bar n\mu} \cP^\perp _\mu -g \tilde\cB_{S\perp}^{n\mu} S_n^\dagger S_{\bar n} g\tilde\cB_{S\perp \mu}^{\bar n}      \nn \\
&\qquad  -\frac{n^\mu \bar n^\nu}{2}   S_n^\dagger ig\tilde G^{\mu \nu}_s S_{\bar n}    \bigg \} ^{BC}    \,,  \nn \\
\cO_s^{q_n B}&=8\pi \alpha_s \left\{  \bar \psi_S^n T^B \frac{\Sl n}{2}  \psi^n_S  \right \} \,, \nn\\
\cO_s^{g_n B}&= 8\pi \alpha_s \left \{   \frac{i}{2} f^{BCD} \cB^{nC}_{S\perp \mu} \frac{n}{2} \sdt (\cP +\cP^\dagger) \cB_{S\perp}^{nD\mu} \right \}\,.
\end{align}
In \Eq{eq:Glauber_Lagrangian}, the operator $\cO_s^{BC}$ connects two operators of different collinear sectors, and describes an arbitrary number of soft gluon emissions from the forward scattering. For zero emissions, it reduces to $8\pi \alpha_s \cP_\perp^2 \delta^{BC}$, which, together with the factors of $1/\cP_\perp^2$ in Eq.~(\ref{eq:Glauber_Lagrangian}), reproduces the expected $1/\cP_\perp^2$ tree level Glauber potential between two collinear partons.
For a single emission, it reduces to the Lipatov vertex~\cite{Kuraev:1976ge}. The Feynman rules for two soft emissions can be found in~\cite{Rothstein:2016bsq}. 

SCET with Glauber operators provides an operator based formalism for studying Glauber exchanges, and the Regge limit of QCD. For example, amplitude level Reggeization and the BFKL equation can be derived in the EFT through the renormalization group evolution of the operators~\cite{Rothstein:2016bsq}. The role of Glauber exchanges for factorization violation can also be explicitly computed within this framework, as discussed in~\cite{Rothstein:2016bsq}. For example, it was used in~\cite{Schwartz:2017nmr} to give direct computations of the collinear factorization violation in spacelike splitting functions that was first found and computed in~\cite{Catani:2011st}. Higher order leading power calculations in the framework used here were also made in \cite{Zhou:2017his}.

\section{Fermionic Glauber Operators}\label{sec:ferm_glaub}

Having reviewed SCET with Glauber gluon operators, in this section we extend the framework to include Glauber quark operators. In \Sec{sec:nn_ops} we describe the structure of the $n-\bn$ scattering operators, and in \Sec{sec:ns_ops} we describe the structure of the $n$-$s$ scattering operators. In \Sec{sec:regs} we discuss the regulators beyond dimensional regularization that are required for calculating with these operators at loop level. The precise structure of the operators presented in this section are derived from the symmetries of the effective theory, power counting and mass dimension constraints, and matching calculations, and are discussed in detail in \Sec{sec:tree_level}.

\subsection{$n$-$\bar n$ Operator Structure}\label{sec:nn_ops}

In this section we present the structure of the $n$-$\bar n$ scattering 
operators that describe the forward scattering of partons in the $n$ and $\bar n$ collinear sectors through the $t$-channel exchange of a Glauber quark. Analogous to the gluon case, in \Eq{eq:Glauber_Lagrangian}, we write the Lagrangian in the factorized form
\begin{align}\label{eq:Glauber_Lagrangian_cquark}
\cL^{\text{II}(1)} \supset e^{-ix\cdot \cP} \sum\limits_{n,\bar n} \bar{\cO}_{\bar n} \frac{1}{\Sl{\cP}_\perp}  \cO_s   \frac{1}{\Sl{\cP}_\perp} \cO_{n}  \,,
\end{align}
where $\cO_{\bar n}$ and $\cO_n$ describe fields in the collinear sectors, while $\cO_s$ describes fields in the soft sector, which sits at an intermediate rapidity between the two collinear sectors. The superscript $\text{II}$ denotes that we are working in SCET$_\text{II}$, and the superscript $(1)$ denotes that this will give contributions that scale as $\cO(\lambda)$. The factors of $\Sl{\cP}_\perp$ indicate that this is a non-local potential, and reflect the fermionic nature of the Glauber quark. We have kept the color and Dirac indices implicit.  To simplify the notation, we will often refer to the operator as
\begin{align} \label{eq:Glauber_nn}
\cO_{\bar n n}=\bar{\cO}_{\bar n} \frac{1}{\Sl{\cP}_\perp}  \cO_s   \frac{1}{\Sl{\cP}_\perp} \cO_{n}\,.
\end{align}
 
In \Eq{eq:Glauber_Lagrangian_cquark}, we have used the $\supset$ notation to emphasize that this is only the component of the subleading Lagrangian, ${\cal L}^{(1)}$, that describes the $t$-channel exchange of a Glauber quark. In particular, it does not describe $\cO(\lambda)$ power corrections to the $t$-channel exchange of a Glauber gluon, or of compound states. In general, there are other operators consistent with the symmetries of the effective theory as well as with power and mass dimension counting that can be written down. For example, in \Eq{eq:Glauber_Lagrangian_cquark}, one may replace $\frac{1}{\Sl{\cP}_\perp}$ with $\frac{1}{\cP_\perp^2}$, and appropriately modify the numerator with an additional derivative or gluon field to satisfy power and mass dimension counting. In \Sec{sec:tree_level} we will show that $\cO_{\bar n n}$ is sufficient for tree level matching, and therefore any additional operators have vanishing Wilson coefficients at this order. Moreover, we find that the one-loop renormalization of $\cO_{\bar n n}$ does not produce additional operators. Hence, \Eq{eq:Glauber_Lagrangian_cquark} is the complete basis of operators for describing quark Reggeization at LL order.  We have not ruled out the presence of additional fermionic exchange operators from one-loop matching, and we leave the study of the general operator basis to future work.

The exchange of a quark necessarily changes the fermion number in each collinear sector. In particular, there are $8$ scattering configurations: 
\begin{align}\label{eq:8configs}
\fd{2.8cm}{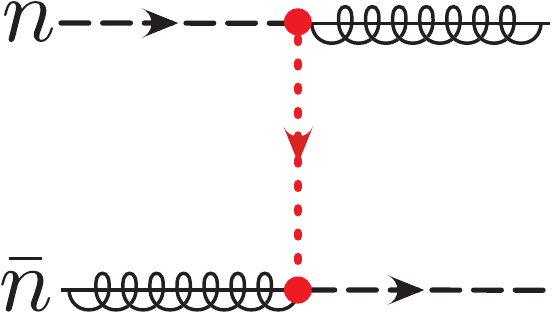} \qquad 
\fd{2.8cm}{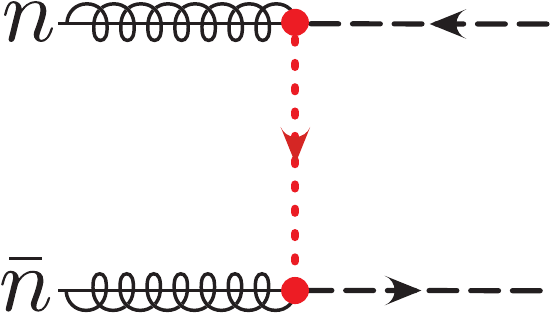}\qquad 
\fd{2.8cm}{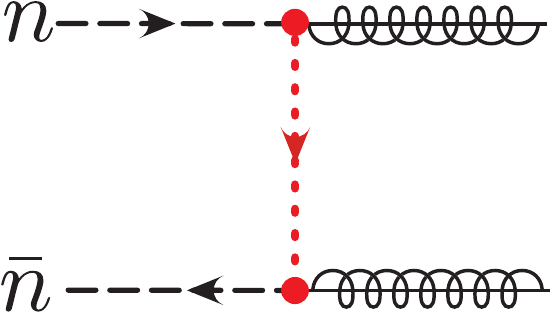} \qquad 
\fd{2.8cm}{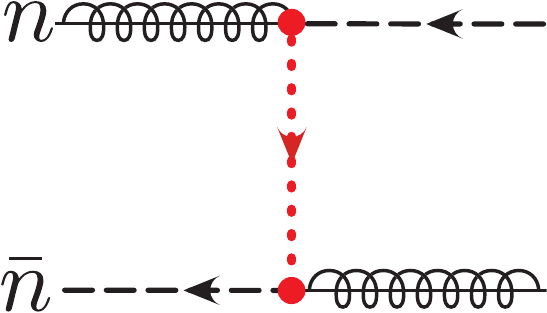} \nn \\[10pt]
\fd{2.8cm}{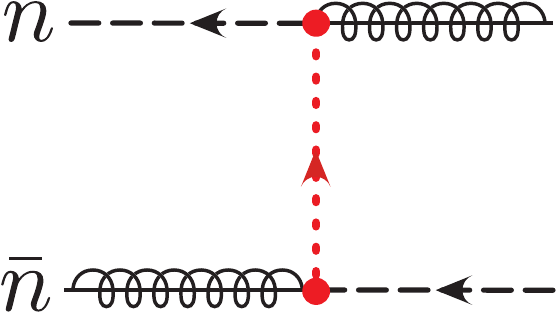} \qquad 
\fd{2.8cm}{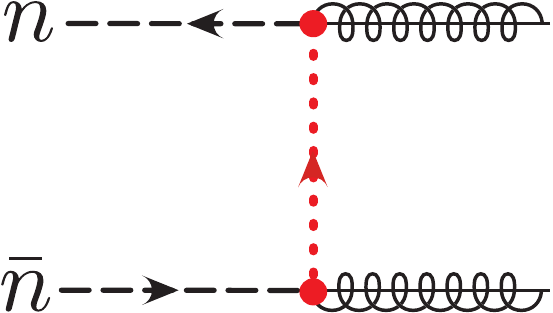}\qquad 
\fd{2.8cm}{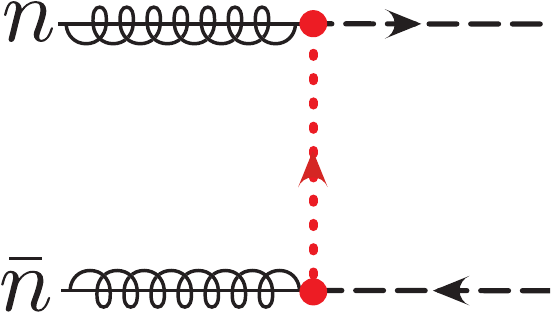} \qquad 
\fd{2.8cm}{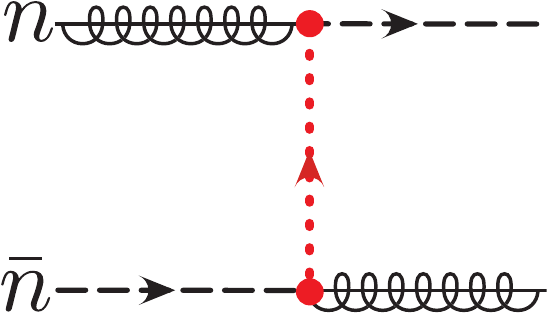}\,,
\end{align}
where the red dotted line denotes the Glauber quark.
Importantly, in \Eq{eq:Glauber_Lagrangian_cquark}, there is a sum over the directions $n$ and $\bar n$, as well as an implicit sum over the label momentum $\cP$. This implies that all $8$ possible collinear-collinear forward scattering configurations are generated from \Eq{eq:Glauber_Lagrangian_cquark}. For scattering configurations that preserve fermion number in each collinear sector, \Eq{eq:Glauber_Lagrangian_cquark} contributes through $T$-products, starting at $\cO(\lambda^2)$ with $T$-products of the above diagrams.

The collinear operators appearing in \Eq{eq:Glauber_nn} are given by
\begin{align}\label{eq:coll_ops}
\cO_{\bar n}=\Sl{\cB}_{\perp \bar n} \chi_{\bar n}\,, \qquad \cO_{n}=\Sl{\cB}_{\perp n} \chi_n\,,
\end{align}
and the soft operator is given by
\begin{equation}\label{eq:soft_operator_quark}
\cO_s = -2\pi \alpha_s \left[ S_{\bar n}^\dagger S_{n} \Sl{\cP}_\perp + \Sl{\cP}_\perp S_{\bar n}^\dagger S_{n} - S_{\bar n}^\dagger S_n g\Sl{\cB}^{n}_{S\perp}-  g\Sl{\cB}^{\bar n}_{S\perp} S_{\bar n}^\dagger S_n \right]\,.
\end{equation}
Note the identity
\begin{align}
\cP^\mu_\perp S_\bn^\dagger S_{n} - S_\bn^\dagger S_{n} g \cB_{S\perp}^{n \mu} =S_\bn^\dagger S_{n} \cP^\mu_\perp -g\cB_{S\perp}^{\bn \mu} S_\bn^\dagger S_{n} \,,
\end{align}
which allows us to write the soft operator in a more compact but less symmetric form. The power counting of the operators is $\cO_n \sim \cO_{\bar n} \sim \lambda^2$ and $\cO_s\sim \lambda$. Using the power counting formula of \cite{Rothstein:2016bsq} which subtracts 2 for a mixed $n$-$\bn$-soft operator, we then find that $\cO_{n\bar n} \sim \lambda$ as stated above.

The structure of the soft operator $\cO_s$ in \Eq{eq:soft_operator_quark} is significantly simpler than for the gluon case, $\cO_s^{BC}$ in \Eq{eq:iain_ira_soft}, due to the difference in mass dimension between fermionic and bosonic propagators.  In the gluon case, $\cO_s$ is exact: it is not corrected at higher orders in perturbation theory since Glauber exchange is instantaneous in both time and longitudinal position, and there is no hard contribution that is integrated out~\cite{Rothstein:2016bsq}. While we expect this to be the case here, due to the possibility of the additional operators mentioned below \Eq{eq:Glauber_nn} appearing at higher orders, and the behavior of power suppressed terms from loop diagrams, it is more complicated to show that this is true in this case, and we leave it to future work.

\begin{figure}
\begin{center}
\begin{align}
\fd{2.8cm}{figures/tree_level_1_low.pdf} &=  \bar u_{\bar n}(p_3) \Sl{\epsilon}_{\!\perp} \! (p_2) T^A \Bigg[  -i g^2 \frac{1}{\Sl{q}_\perp}  \Bigg]  \Sl{\epsilon}_{\!\perp} \!(p_4) T^B u_n(p_1) \nn \\[10pt]
\fd{2.8cm}{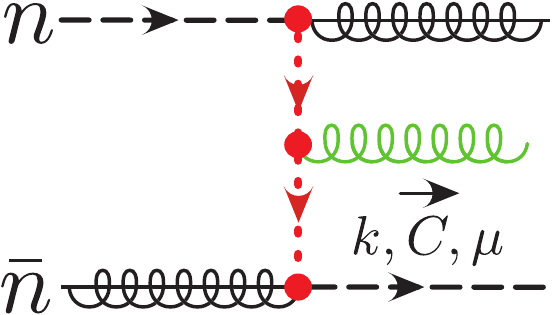} &=\bar u_{\bar n}(p_3) \Sl{\epsilon}_{\!\perp} \! (p_2) T^A \Bigg[   i g^3~ T^C \frac{1}{\Sl{q}_\perp}  \bigg( \gamma^\mu_\perp - \frac{(\Sl{q}_\perp+\Sl{k}_\perp) n^\mu}{n\cdot k}      \nn \\
&\ + \frac{\Sl{q}_\perp  \bar n^\mu}{\bar n \cdot k} \bigg) \frac{1}{\Sl{q}_\perp +  \Sl{k}_\perp}   \Bigg]  \Sl{\epsilon}_{\!\perp} \!(p_4) T^B u_n(p_1)  \nn \\[10pt]
\fd{3.6cm}{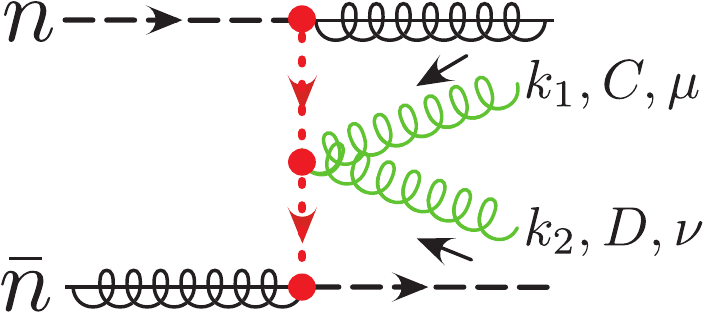}  &= \bar u_{\bar n}(p_3) \Sl{\epsilon}_{\!\perp} \! (p_2) T^A \Bigg[ -i g^4 T^C T^D \frac{1}{\Sl{q}_{\perp} +\Sl{k}_{1\perp} + \Sl{k}_{2\perp}} \bigg( \frac{ n^\nu \gamma^\mu_{\perp} }{ n \cdot k_2} 
-\frac{ \bn^\mu \gamma^\nu_{\perp}}{ \bn \cdot k_1}   \nn \\
&\  + \frac{(\Sl{q}_{\perp} +\Sl{k}_{1\perp} + \Sl{k}_{2\perp})  \ \bn^\mu   \bn^\nu  }{2\bn \cdot( k_1+ k_2) \bn \cdot k_1 } + \frac{\Sl{q}_\perp n^\mu  n^\nu }{2n \cdot( k_1+ k_2) n \cdot k_1}   \nn \\
&\ 
- \frac{( \Sl{k}_{1\perp } +  \Sl{q}_\perp) \bn^\mu n^\nu }{\bn\cdot k_1\,n \cdot k_2} \bigg) \frac{1}{\Sl{q}_\perp} + \Big\{ (C, \mu, k_1) \leftrightarrow (D, \nu, k_2) \Big\}  \Bigg]  \Sl{\epsilon}_{\!\perp} \!(p_4) T^B u_n(p_1) 
 \nn
\end{align}
\end{center}
\caption{Feynman rules for tree level $qg$ forward scattering with zero, one and two soft gluon emissions, generated by the soft operator $\cO_s$.  Soft emissions at higher orders in $\alpha_s$ are also produced by $\cO_s$.
}\label{fig:no_emission_feynrule}
\end{figure}

The soft operator $\cO_s$ describes the emission of soft gluons from the forward scattering to all orders in $\alpha_s$.
The Feynman rules for $qg$ forward scattering with zero, one and two soft gluon emissions are given in \Fig{fig:no_emission_feynrule}. The one emission Feynman rule gives the classic result of Fadin and Sherman \cite{Fadin:1976nw,Fadin:1977jr}, which we will refer to as the Fadin-Sherman vertex. The two emission Feynman rule has not, to our knowledge, appeared in the literature before. It will be required in our derivation of the quark Reggeization through rapidity renormalization (although only a particularly simple projection appears).

\subsection{$n$-$s$ Operator Structure}\label{sec:ns_ops}
In addition to the $n$-$\bar n$ scattering operators, the effective theory also includes operators that describe $n$-$s$ (and $\bn-s$) forward scattering.  We write the Lagrangian for soft collinear forward scattering as
\begin{align}\label{eq:Glauber_Lagrangian_squark}
\cL_G^{\text{II}(1/2)} \supset e^{-ix\cdot \cP} \sum\limits_{n}  \bar{\cO}_n \frac{1}{\Sl{\cP}_\perp}  \cO^{n}_s  + \bar{\cO}^{n}_s \frac{1}{\Sl{\cP}_\perp} \cO_n  \,.
\end{align}
Here the superscript $1/2$ indicates that this Lagrangian contribution scales as $\cO(\lambda^{1/2})$ relative to the leading power contribution. 
These operators play an important role in the rapidity renormalization, contributing through $T$-products in the effective theory. In particular, their contribution scales as $\cO(\lambda^{1/2}) \cdot \cO(\lambda^{1/2})= \cO(\lambda)$, which is at the same order as the $n$-$\bar n$ forward scattering operators. We will use the shorthand
\begin{align}\label{eq:Glauber_ns}
\cO_{ns}=\bar{\cO}_n \frac{1}{\Sl{\cP}_\perp}  \cO^{n}_s  
 \,.
\end{align}

As in \Eq{eq:Glauber_Lagrangian_cquark}, we have used the $\supset$ symbol in \Eq{eq:Glauber_Lagrangian_squark} to emphasize that this is not the complete Lagrangian at $\cO(\lambda^{1/2})$, and includes only the operators required for describing quark Reggeization at LL order

In \Eq{eq:Glauber_Lagrangian_squark}, the sum over the direction $n$, the implicit sum over the label momentum $\cP$, and the presence of both $\cO_{ns}$ and its hermitian conjugate generates all possible scattering configurations, namely:
\begin{align}
\fd{2.8cm}{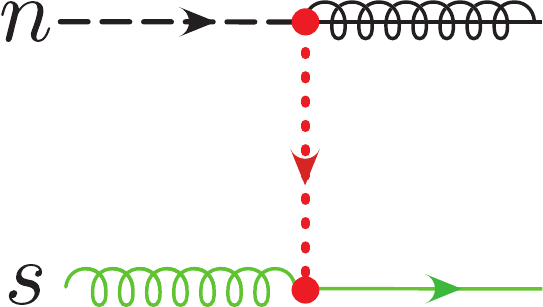} \qquad 
\fd{2.8cm}{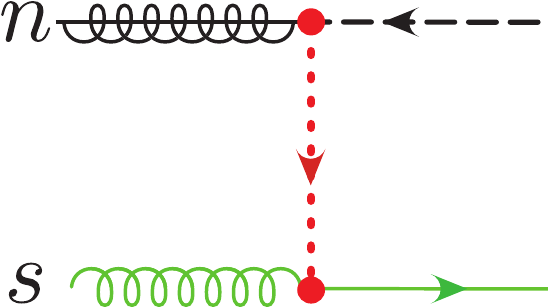}\qquad 
\fd{2.8cm}{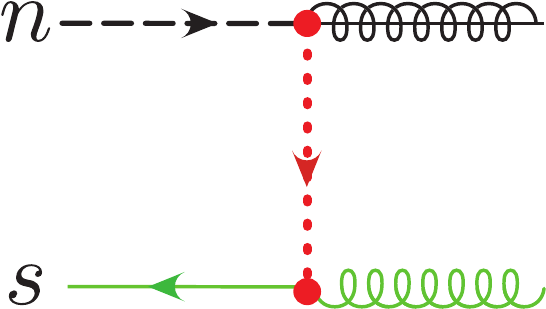} \qquad 
\fd{2.8cm}{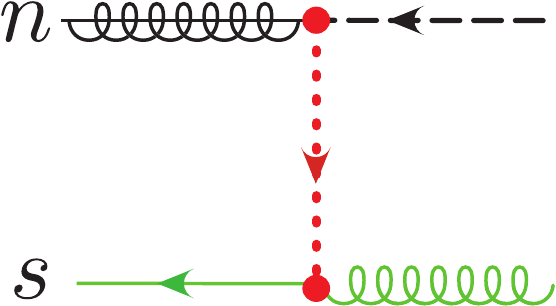}\nn 
\phantom{\,.} \\[10pt]
\fd{2.8cm}{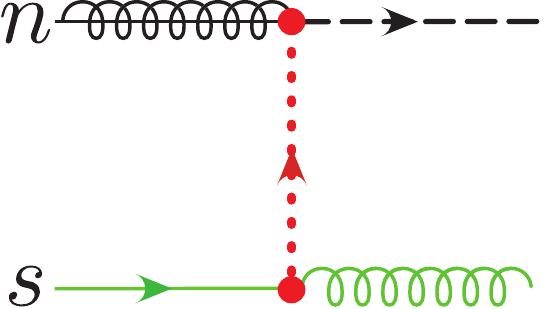} \qquad 
\fd{2.8cm}{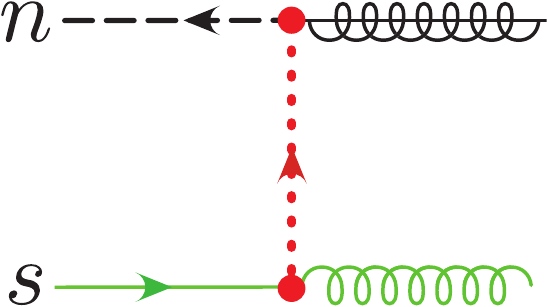}\qquad 
\fd{2.8cm}{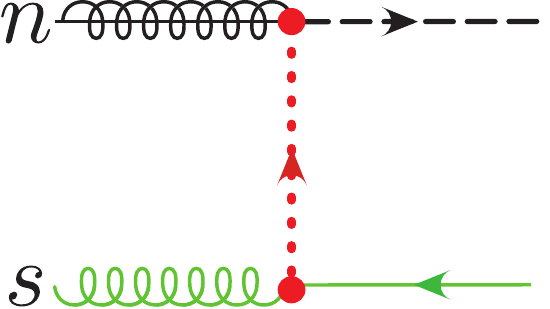} \qquad 
\fd{2.8cm}{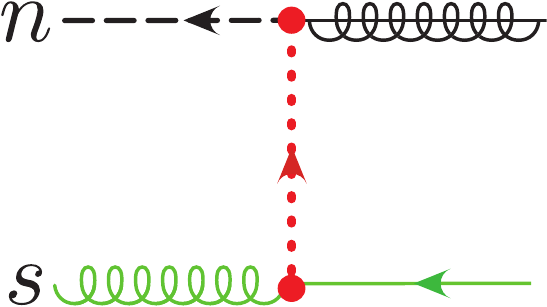}\,.
\end{align}

The $\cO_n$ operators in \Eq{eq:Glauber_Lagrangian_squark} are identical to those in \Eq{eq:coll_ops}. The $\cO_s^{n}$ operators have a similar structure, but we include a prefactor that arises from tree level matching:
\begin{align}\label{eq:n_s_ops}
\cO^{n}_s=-4\pi \alpha_s \Sl{\cB}_{\perp S}^n \psi^n_S\,, 
\qquad \bar{\cO}^n_s=-4\pi \alpha_s \bar \psi^n_S \Sl{\cB}_{\perp S}^n \,.
\end{align}
The power counting of the operators is $\cO_n \sim \cO_{\bar n} \sim \lambda^2$ and $\cO_s^n \sim\cO_s^{\bar n} \sim \lambda^{3/2}$. Using the power counting formula of \cite{Rothstein:2016bsq}, where we subtract 3 for a mixed $n$-soft or $\bn$-soft operator, we then find that $\cO_{ns} \sim \lambda^{1/2}$, as stated.

\subsection{Regulators for Rapidity and Glauber Potential Singularities}\label{sec:regs}

As discussed extensively in \cite{Rothstein:2016bsq}, the Glauber Lagrangian requires both the regularization of rapidity divergences, as well as the regularization of divergences associated with Glauber exchanges. Here we use identical regulators to those defined in  \cite{Rothstein:2016bsq}.

Rapidity divergences are regulated using the $\eta$-regulator of \cite{Chiu:2012ir,Chiu:2011qc}. In this regulator the soft and collinear Wilson lines are modified as
\begin{align}
S_n&= \left[  \sum\limits_{\text{perms}} \exp \left(  -\frac{g}{n\cdot \cP} \frac{\omega |2\cP^z|^{-\eta/2}}{\nu^{-\eta/2}} n \cdot A_s(x)  \right) \right]\,, \nn \\
W_{n}&= \left[  \sum\limits_{\text{perms}} \exp \left(  -\frac{g}{ \bar n\cdot \cP} \frac{\omega^2 |\bar n\cdot \cP|^{-\eta}}{\nu^{-\eta}}  \bar n \cdot A_{n}(x)  \right) \right]\,,
\end{align}
with analogous modifications for $S_{\bar n}$ and $W_{\bar n}$. Here $\omega$ is a formal bookkeeping parameter which satisfies
\begin{align}
\nu \frac{\partial}{\partial \nu} \omega^2(\nu) =-\eta~ \omega^2(\nu) \,, \qquad \lim_{\eta \to 0}\, \omega(\nu)=1\,.
\end{align}
For convenience we set $\omega=1$ throughout our calculations since it can be trivially restored. 

Singularities from Glauber exchanges are also regulated using the $\eta$-regulator. In particular, a factor of $\omega |2q^z|^{-\eta} \nu^\eta$ is included for each Glauber exchange, where $q$ is the Glauber momentum. This can be formulated at the level of the Glauber Lagrangian, and can be shown to be routing independent \cite{Rothstein:2016bsq}. We regulate divergences associated with Glauber quarks in an identical manner, and show the consistency of this regulator at one-loop through our calculations of the Reggeization, the BFKL equation, and the box diagrams with simultaneous exchange of a Glauber quark and a Glauber gluon.

\section{Tree Level Matching}\label{sec:tree_level}

In this section we consider tree level matching between QCD and SCET. This, combined with the symmetries of the effective theory as well as constraints from power and mass dimension counting, will allow us to fix the structure of the operators, as given in the previous section. In \Sec{sec:nbarn_match} and \Sec{sec:ns_match} we perform the matching with zero soft emissions. In \Sec{sec:soft_match} we present the most general form of the soft operator $\cO_s$, and fix its structure with tree level matching.

We will use the following alternative notation for Feynman diagrams involving Glauber quark exchange, distinguishing the Glauber quark exchange from a Glauber gluon exchange by including an arrow on the red dotted line:
\begin{align}\label{eq:collinear_potential}
\fd{2.8cm}{figures/tree_level_1_low.pdf}\equiv\fd{2.6cm}{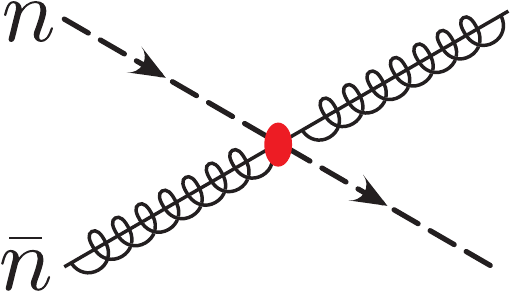}\,, \qquad 
\fd{2.8cm}{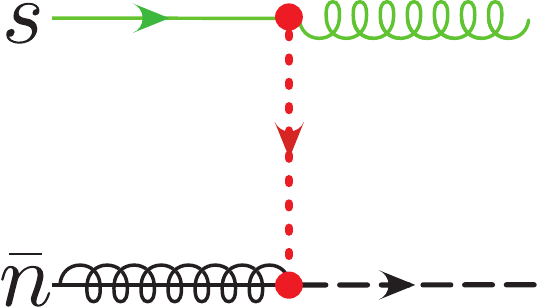}\equiv\fd{2.3cm}{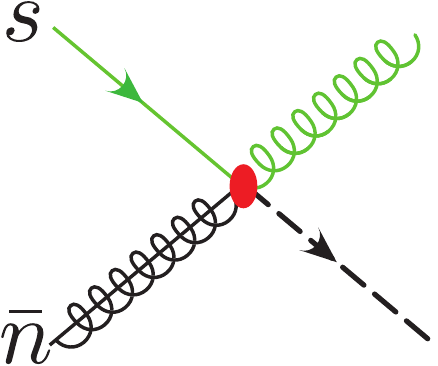}\,,
\end{align}
where we have illustrated with particular configurations of $n$-$\bar n$ and $n$-$s$ scattering. The notation with the red dotted line shows the $t$-channel exchange explicitly, while the notation with the red elliptical blob emphasizes the potential nature of the forward scattering operators.

\subsection{$n$-$\bar n$ Scattering}\label{sec:nbarn_match}

We begin with the matching for the $n$-$\bar n$ scattering operator. For definiteness, we take the configuration $q(p_1^n) + g(p_2^{\bn}) \to  g(p_4^n) + q(p_3^{\bn})$, and choose our momenta as 
\begin{align}
p_{1\perp}&=-p_{4\perp}=q_\perp/2 \,, \qquad p_{2\perp}=-p_{3\perp}=-q_\perp/2 \,.
\end{align}
For this choice, the positive $q_\perp$ is aligned with the fermion number flow.
Expanding the full theory result in the forward limit, we find
\begin{align}\label{eq:full_theory_zero}
\fd{3.5cm}{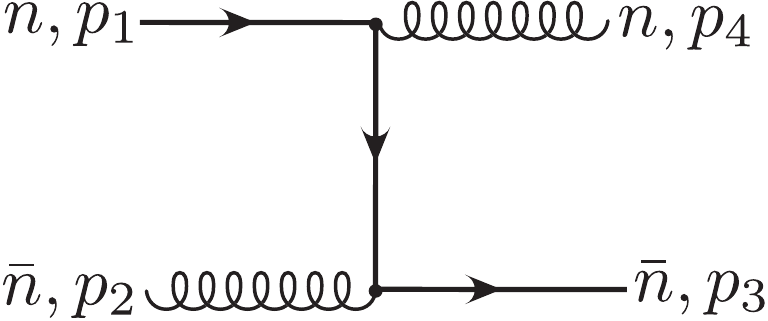}=-4\pi i \alpha_s \bar u_{\bar n}(p_3) \Sl{\epsilon}_\perp (p_2) T^A \frac{\Sl{q}_\perp}{q_\perp^2}  \Sl{\epsilon}_\perp (p_4) T^B u_n(p_1)\,.
\end{align}
This is reproduced in the effective theory by the zero emission Feynman rule of the forward scattering operator $\cO_{\bn n}$:
\begin{align}
\fd{2.6cm}{figures/fermion_regge_vertex.pdf} =\langle \cO_{\bn n} \rangle=\left \langle \bar \chi_{\bar n} \Sl{\cB}_{\perp \bar n} \frac{1}{\Sl{\cP}_\perp} (-4\pi \alpha_s \Sl{\cP}_\perp) \frac{1}{\Sl{\cP}_\perp} \Sl{\cB}_{\perp n} \chi_n \right \rangle\,.
\end{align}
In particular, this defines the normalization of the soft operator $\cO_s$ with zero emissions, but does not probe the structure of the soft Wilson lines or the soft gluon fields within $\cO_s$.

\subsection{$n$-$s$ Scattering}\label{sec:ns_match}

The expansion of the full theory diagram in \Eq{eq:full_theory_zero} also fixes the structure of the $n$-$s$ operators. In particular, we immediately see that it is reproduced by the zero emission Feynman rule of the forward scattering operator $\cO_{\bn s}$:
\begin{align}
\fd{2.3cm}{figures/soft_scatter_regge_vertex_low.pdf}= \langle \cO_{\bn s} \rangle = \left \langle\bar \chi_{\bar n} \Sl{\cB}_{\perp \bar n} \frac{1}{\Sl{\cP}_\perp}  \left( -4\pi \alpha_s \Sl{\cB}_{\perp S}^\bn \psi^\bn_S \right)\right\rangle\,.
\end{align}
This simple matching, combined with constraints from power counting, mass dimension and the symmetries of the effective theory, therefore fixes the form of the operators $\cO_{ns}$ and $\cO_{\bn s}$. Once again these are the only operators that appear from tree level matching.

\subsection{Matching to the Soft Operator}\label{sec:soft_match}

To derive the precise structure of the soft operator $\cO_s$, we must consider matching with soft gluon emissions. We begin by deriving the most general form of the soft operator consistent with constraints from power counting, mass dimension and the symmetries of the effective theory. We then use matching calculations to fix the free coefficients in the operator.

The soft operator must have mass dimension $1$, scale as $\cO(\lambda)$, and be composed of gauge invariant building blocks in the effective theory such as $\cP_\perp$, $\cB^{\bar n \mu}$ and Wilson lines. Since the total $\perp$ momentum of the Lagrangian is zero, we have $\cP_\perp=\cP_\perp^\dagger$, and therefore we can choose to write the operator in terms of $\cP_\perp$. Hermiticity requires that the operator satisfies (up to $\gamma_0$ factors that are absorbed by the collinear operators in $\cO_{\bn n}$)
\begin{align}
\cO_s=\cO^\dagger_s \big|_{n \leftrightarrow \bar n} \,.
\end{align}
The above constraints do not prohibit the appearance of an arbitrary number of soft Wilson lines since these have mass dimension $0$ and scale as $\cO(\lambda^0)$. However, due to the physical picture of these Wilson lines as arising from the emission of gluons off the partons involved in the forward scattering, we will require that each term in the soft operator has two Wilson lines. These soft Wilson lines can appear both explicitly, as well as inside the gauge invariant soft gluon fields, defined in \Eq{eq:gauge_soft}, and both must be counted. The constraint of having two soft Wilson lines leads to the following allowed combinations:
\begin{align}
S_{\bar n}^\dagger S_n \cB^{n\mu}_{S\perp}\,, \qquad \cB^{\bar n \mu}_{S\perp} S_{\bar n}^\dagger S_n\,, \qquad \cB^{n\mu}_{S\perp} S_n^\dagger S_{\bar n} \,, \qquad S_n^\dagger S_{\bar n} \cB^{\bar n \mu}_{S\perp}\,.
\end{align}

Given these constraints, the most general structure of the operator is
\begin{align}\label{eq:gen_soft}
\cO_s &= -4\pi \alpha_s \left[ \frac{C_1}{2} \left(g\Sl{\cB}^{n}_{\perp s}  S_n^\dagger S_{\bar n} + S_n^\dagger S_{\bar n} g\Sl{\cB}_{\perp s}^{\bar n} \right) +\frac{C_2}{2} \left(S_{\bar n}^\dagger S_n g\Sl{\cB}^{n}_{S\perp} +  g\Sl{\cB}^{\bar n}_{S\perp} S_{\bar n}^\dagger S_n \right) \right. \nn \\
&+\frac{C_3}{2} \left( S_n^\dagger S_{\bar n} \Sl{\cP}_\perp + \Sl{\cP}_\perp S_n^\dagger S_{\bar n} \right) \left. +\frac{C_4}{2}\left( S_{\bar n}^\dagger S_{n} \Sl{\cP}_\perp + \Sl{\cP}_\perp S_{\bar n}^\dagger S_{n}\right) \right ] \,.
\end{align}
The tree level matching with zero emission in \Sec{sec:nbarn_match} gives the relation
\begin{align}\label{eq:zeromatchingrelation}
C_3+C_4=1\,.
\end{align}
In the next section, we derive additional coefficient relations by considering soft emissions, which probe the structure of the soft Wilson lines and the soft gluon fields. Note that the general form of the soft operator in \Eq{eq:gen_soft} includes both combinations $S_n^\dagger S_{\bar n}$ and $S_{\bar n}^\dagger S_{n}$. In the Glauber gluon case, the soft operator $\cO_s^{BC}$ in \Eq{eq:iain_ira_soft} has only one of these combinations, corresponding to the ordering of the operators $\cO_n^{iB}$, $\cO_s^{BC}$ and $\cO_\bn^{jC}$ in \Eq{eq:Glauber_Lagrangian}. We will see that this also holds in the Glauber quark case, and in particular we will show that $C_1=C_3=0$ for the ordering of operators in \Eq{eq:Glauber_Lagrangian_cquark}.

\subsubsection{One Soft Emission}\label{sec:one_soft}

\begin{figure}[t!]
	\begin{center}
		\raisebox{2.5cm}{
			\hspace{-14.4cm}
			a)
		} \\[-83pt]
		\hspace{-0.5cm}
		\raisebox{0.32cm}{
			\includegraphics[width=0.16\columnwidth]{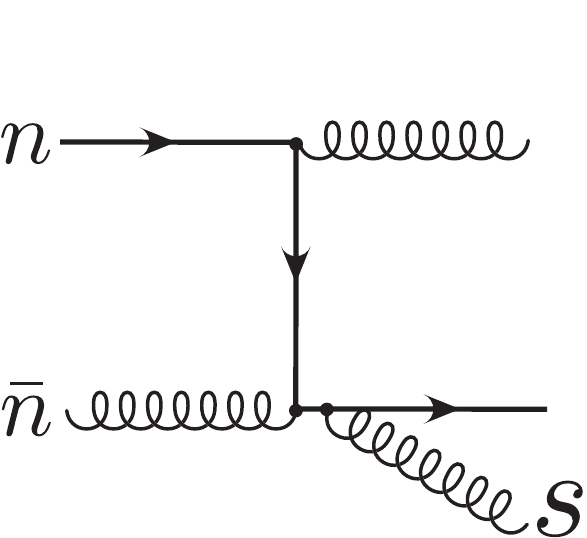}
		}\hspace{0.2cm}
		\raisebox{0.3cm}{
			\includegraphics[width=0.15\columnwidth]{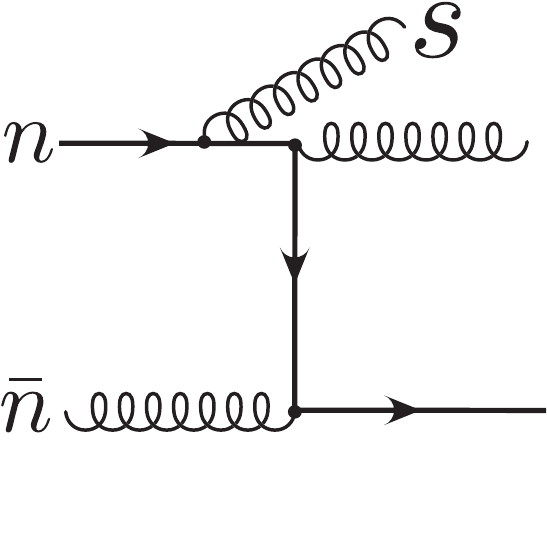} 
		}\hspace{0.2cm}
		\raisebox{0.3cm}{
			\includegraphics[width=0.16\columnwidth]{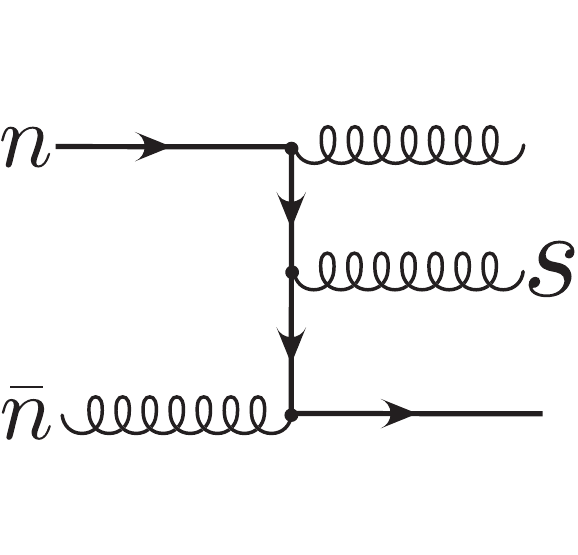}
		}\hspace{0.2cm}
		\raisebox{0.3cm}{
			\includegraphics[width=0.18\columnwidth]{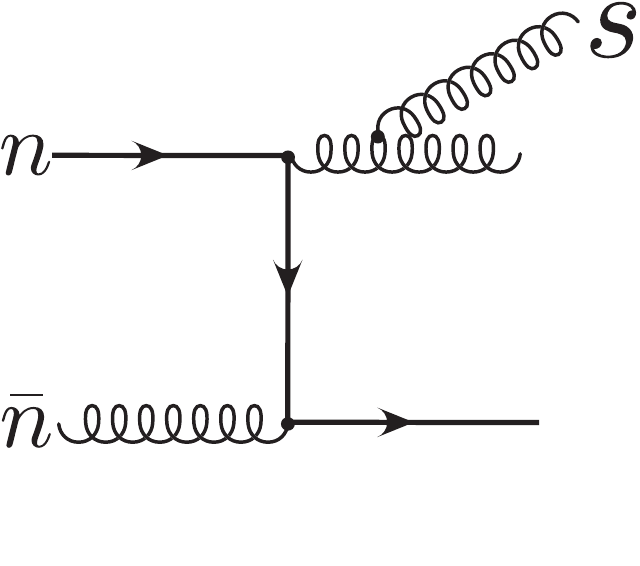}
		}\hspace{0.2cm}		
		\raisebox{0.3cm}{
			\includegraphics[width=0.16\columnwidth]{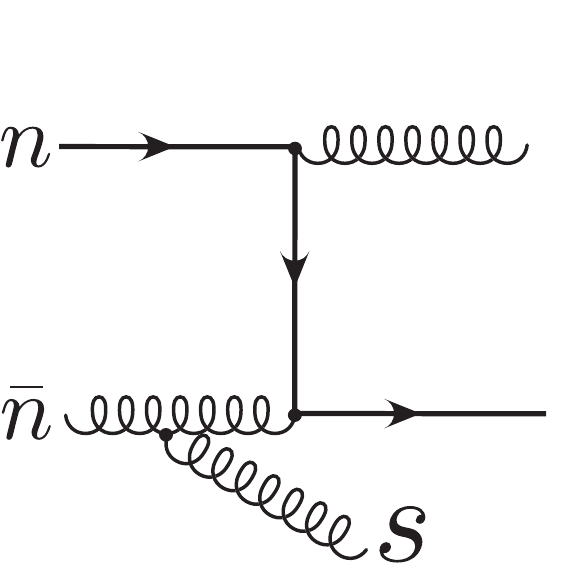}
		}\hspace{0.2cm}
		\\[8pt]
				\raisebox{2cm}{
			\hspace{-1.5cm}
			b)\hspace{3.1cm} 
		} \\[-50pt]	
		\hspace{-0.85cm}
		\includegraphics[width=0.18\columnwidth]{figures/fermion_lipatov_1emission_low.pdf}				
	\end{center}
	\vspace{-0.4cm}
	\caption{\setcaptionskip
	(a) Full theory and  (b) effective theory graphs with a single soft emission. We refer to the effective theory vertex as the Fadin-Sherman vertex since it first appeared in \cite{Fadin:1976nw,Fadin:1977jr}.}
	\label{fig:single_emission}
	\setmainskip
\end{figure}

The single emission diagrams in the full theory and effective theory are shown in \Fig{fig:single_emission}.
Expanded to a single emission with outgoing momentum $k$, the soft operator is given by
\begin{align}\label{eq:soft_op_1emission}
 \cO_s &=-4\pi \alpha_s \left[ \vphantom{\frac{C_3}{2}}   (C_1+C_2)g \Sl{A}_{s\perp} \right. \nn \\
  &-\left(  \frac{C_1}{2}+ \frac{C_2}{2} \right) \left( \frac{gT^A n\cdot A^A_{sk}}{n\cdot k}+\frac{gT^A \bar n\cdot A^A_{sk}}{\bar n \cdot k}  \right) (\Sl{q}_\perp +\Sl{k}_\perp) \nn \\
  &-\left(  \frac{C_3}{2}- \frac{C_4}{2} \right) \left( \frac{gT^A n\cdot A^A_{sk}}{n\cdot k}-\frac{gT^A \bar n\cdot A^A_{sk}}{\bar n \cdot k}  \right) (\Sl{q}_\perp +\Sl{k}_\perp) \nn \\
  &+\left(  \frac{C_1}{2}+ \frac{C_2}{2} \right) \Sl{q}_\perp \left( \frac{gT^A n\cdot A^A_{sk}}{n\cdot k}+\frac{gT^A \bar n\cdot A^A_{sk}}{\bar n \cdot k}  \right) \nn \\
  &\left. -\left(  \frac{C_3}{2}- \frac{C_4}{2} \right)  \Sl{q}_\perp \left( \frac{gT^A n\cdot A^A_{sk}}{n\cdot k}-\frac{gT^A \bar n\cdot A^A_{sk}}{\bar n \cdot k}  \right) \right ]\,.
\end{align}
To fix $C_1+C_2$, we only need the perpendicular polarization, which comes from the full theory diagram
\begin{align}
\fd{2.8cm}{figures/tree_level_full_theory_1emission_3.pdf} \
=i 4\pi \alpha_s  \bar u_{\bar n} \Sl{\epsilon}_{\perp} T^A \frac{\Sl{q}_\perp}{q_\perp^2}  \gamma^\rho_\perp T^c  \frac{(\Sl{q}_\perp  +\Sl{k}_\perp )}{(q_\perp+k_\perp)^2} \Sl{\epsilon}_{\perp} T^B u_n\,.
\end{align}
In the effective theory, we have
\begin{align}
\fd{2.8cm}{figures/fermion_lipatov_1emission_low.pdf}=-i4\pi \alpha_s  (C_1+C_2) \bar u_{\bar n} \Sl{\epsilon}_{\perp} T^A \frac{\Sl{q}_\perp}{q_\perp^2}  \gamma^\rho_\perp T^c  \frac{(\Sl{q}_\perp  +\Sl{k}_\perp )}{(q_\perp+k_\perp)^2} \Sl{\epsilon}_{\perp} T^B u_n\,,
\end{align}
and thus the constraint from matching is
\begin{align}\label{eq:onematchingrelationperp}
C_1+C_2=-1\,.
\end{align}
The Wilson line structure is probed using the $n\cdot A$ and $\bar n \cdot A$ polarizations of the emission. From the remaining four diagrams in the full theory, we find
\begin{align}
&\fd{3cm}{figures/tree_level_full_theory_1emission_1.pdf}+
\fd{2.8cm}{figures/tree_level_full_theory_1emission_2.pdf}+
\fd{3.3cm}{figures/tree_level_full_theory_1emission_4.pdf}+
\fd{3cm}{figures/tree_level_full_theory_1emission_5.pdf}\nn \\[5pt]
&=-i 4\pi  \alpha_s \bar u_{\bar n} \Sl{\epsilon}_{\perp} T^A \left[  \left( \frac{gT^A n\cdot A^A_{sk}}{n\cdot k}  \right) (\Sl{q}_\perp +\Sl{k}_\perp)
  - \Sl{q}_\perp \left( \frac{gT^A \bar n\cdot A^A_{sk}}{\bar n \cdot k}  \right) \right ] \Sl{\epsilon}_{\perp} T^B u_n \,.
\end{align}
Upon comparing with \Eq{eq:soft_op_1emission}, we derive the relation
\begin{align}\label{eq:onematchingrelation}
C_1+C_2&=(C_3-C_4)\,. 
\end{align}

The constraints derived from zero and one emission matching, given in Eqs.~(\ref{eq:zeromatchingrelation}),~(\ref{eq:onematchingrelationperp}) and~(\ref{eq:onematchingrelation}), have the solution $C_1+C_2=-1$, $C_3=0$ and $C_4=1$. The remaining degeneracy between the coefficients $C_1$ and $C_2$ can be broken by matching with two soft emissions.

\subsubsection{Two Soft Emissions}\label{eq:two_soft}

\begin{figure}[t!]
	\begin{center}
		\raisebox{2cm}{
			\hspace{-14.4cm}
			a)
		} \\[-83pt]
		\hspace{-0.5cm}
		\raisebox{0.35cm}{
			\includegraphics[width=0.16\columnwidth]{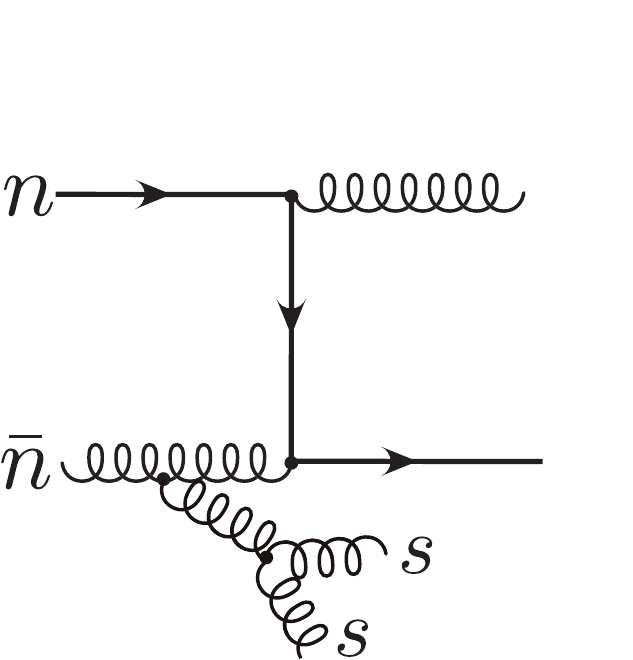}
		}\hspace{0.2cm}
		\raisebox{0.3cm}{
			\includegraphics[width=0.16\columnwidth]{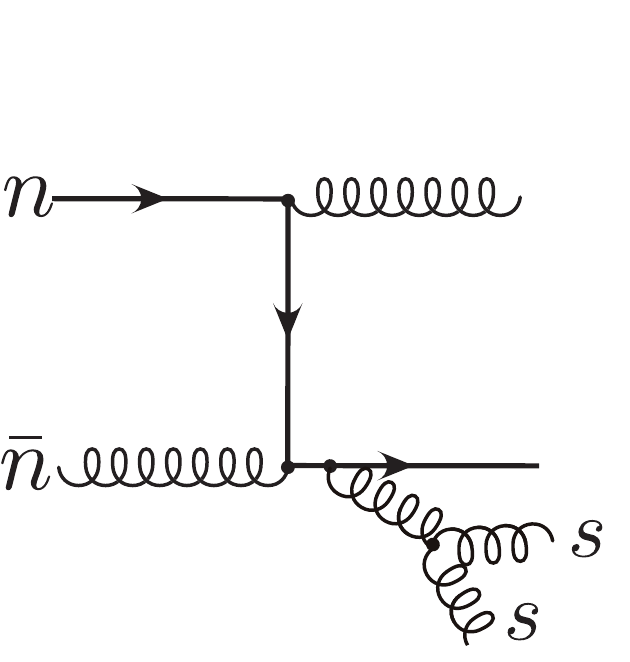} 
		}\hspace{0.2cm}
		\raisebox{0.3cm}{
			\includegraphics[width=0.16\columnwidth]{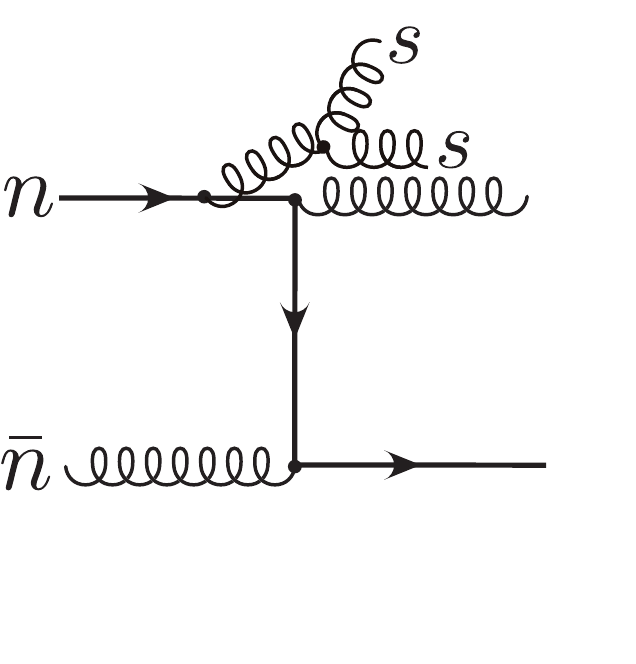}
		}\hspace{0.2cm}
		\raisebox{0.3cm}{
			\includegraphics[width=0.16\columnwidth]{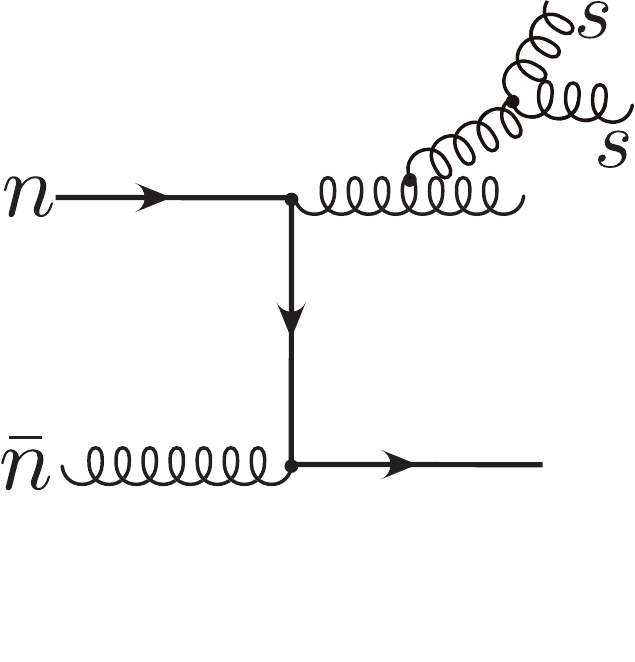}
		}\hspace{0.2cm}
		\\[8pt]
		\hspace{-0.55cm}
		\includegraphics[width=0.16\columnwidth]{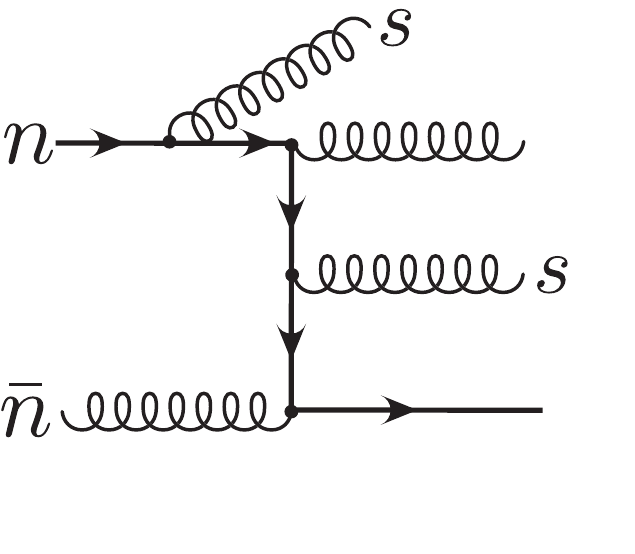}
		\hspace{0.2cm}
		\includegraphics[width=0.16\columnwidth]{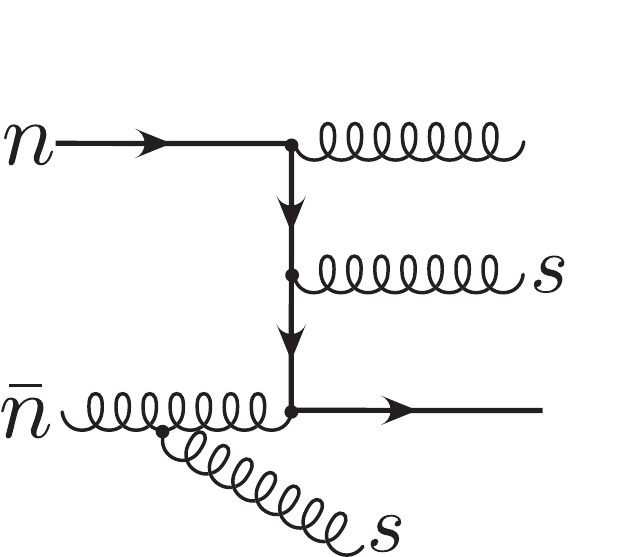}
		\hspace{0.2cm}
		\includegraphics[width=0.16\columnwidth]{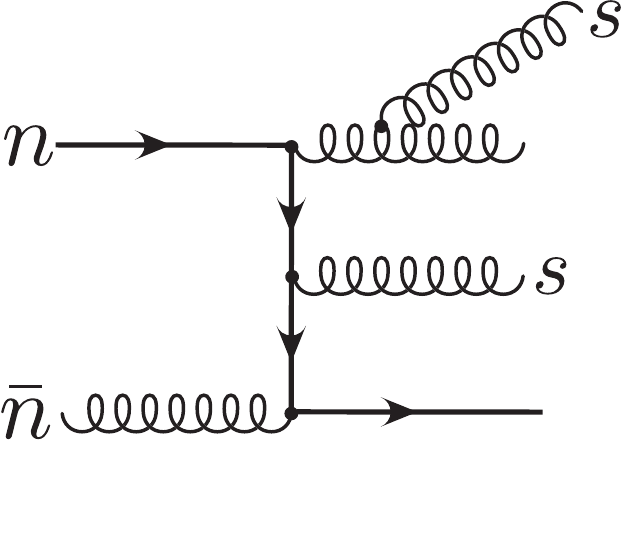}				
		\hspace{0.2cm}
		\includegraphics[width=0.16\columnwidth]{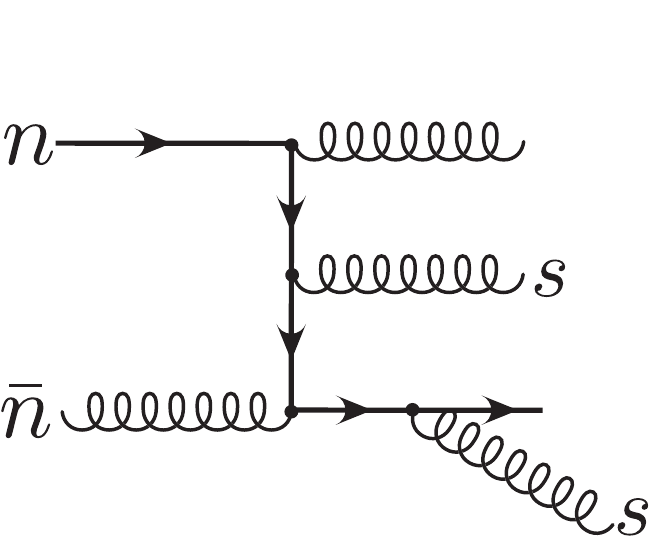}							
		\\[8pt]
		\hspace{-0.55cm}
		\raisebox{-0.1cm}{\includegraphics[width=0.13\columnwidth]{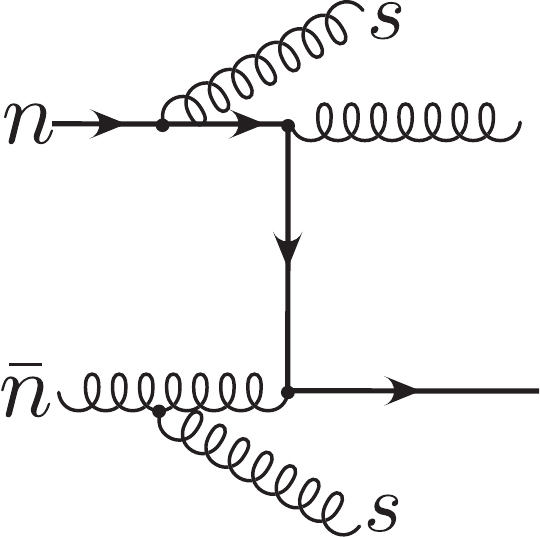}}
		\hspace{0.2cm}
		\raisebox{0.0cm}{\includegraphics[width=0.15\columnwidth]{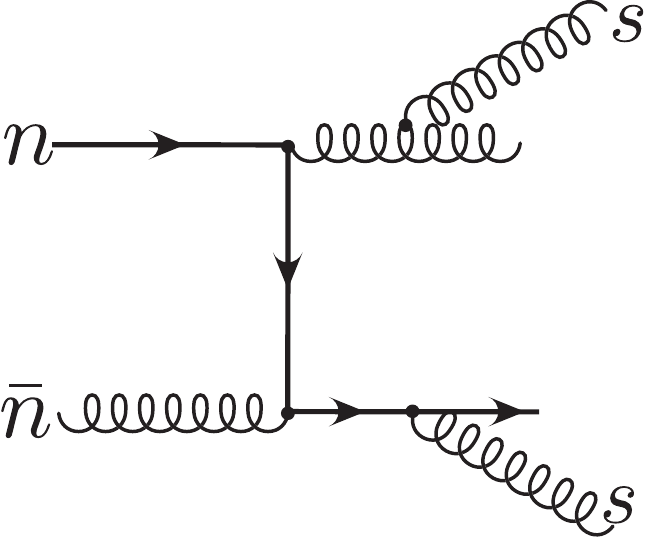}}
		\hspace{0.2cm}
		\raisebox{0.0cm}{\includegraphics[width=0.15\columnwidth]{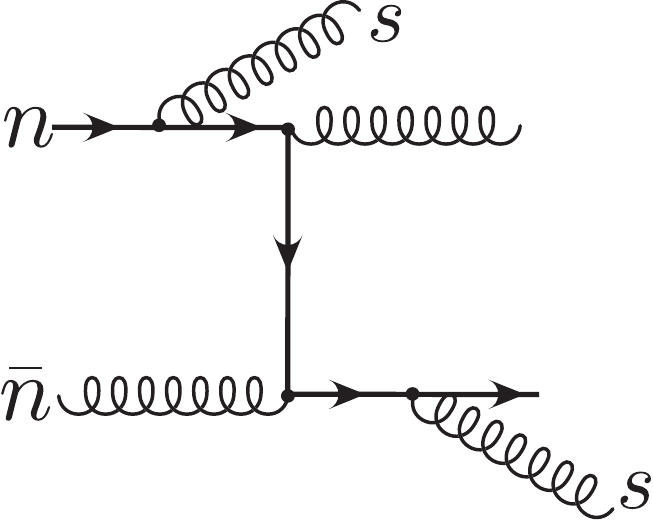}}				
		\hspace{0.2cm}
		\raisebox{-0.1cm}{\includegraphics[width=0.15\columnwidth]{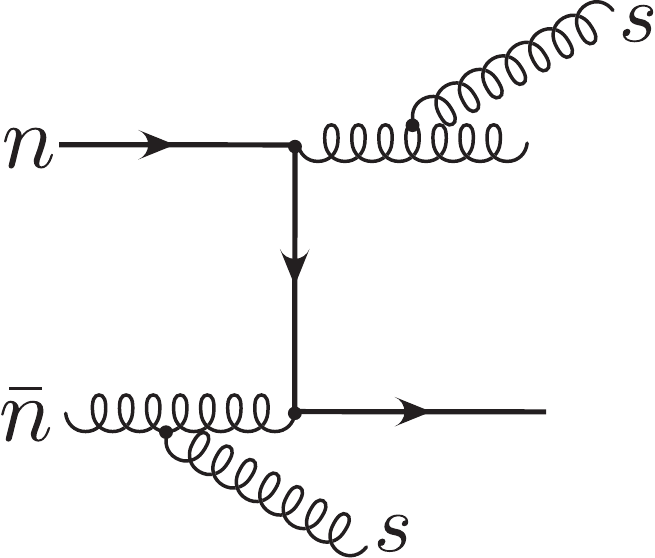}}							
		\\[8pt]
		\hspace{-0.85cm}
		\includegraphics[width=0.15\columnwidth]{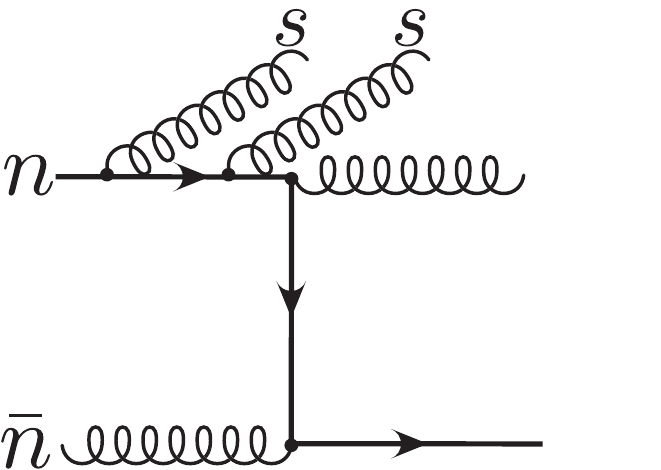}
		\hspace{0.2cm}
		\includegraphics[width=0.15\columnwidth]{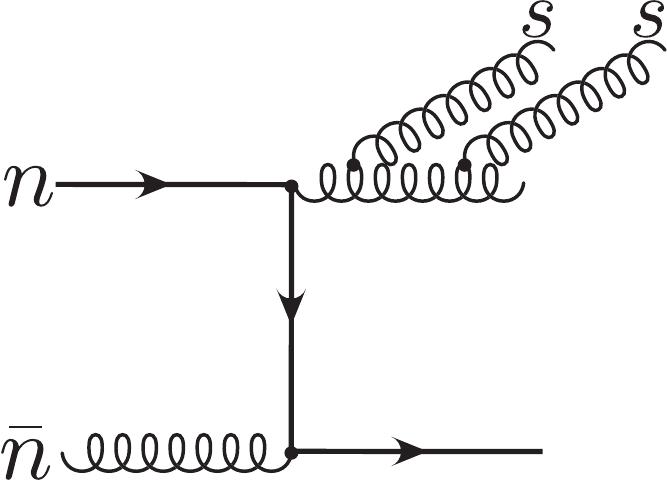}
		\hspace{0.2cm}
		\includegraphics[width=0.15\columnwidth]{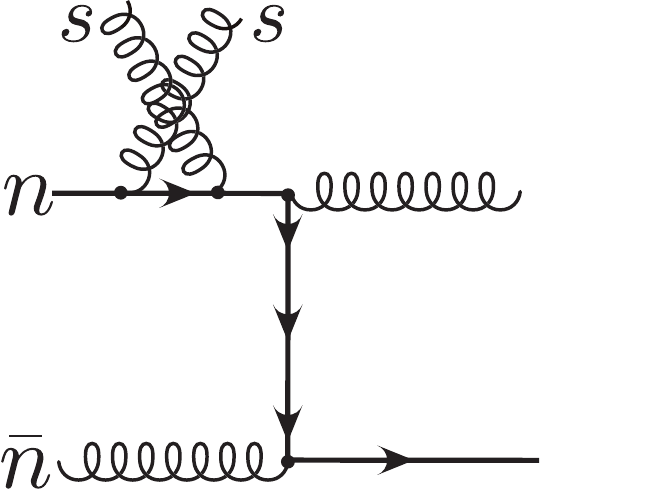}				
		\hspace{0.2cm}
		\includegraphics[width=0.15\columnwidth]{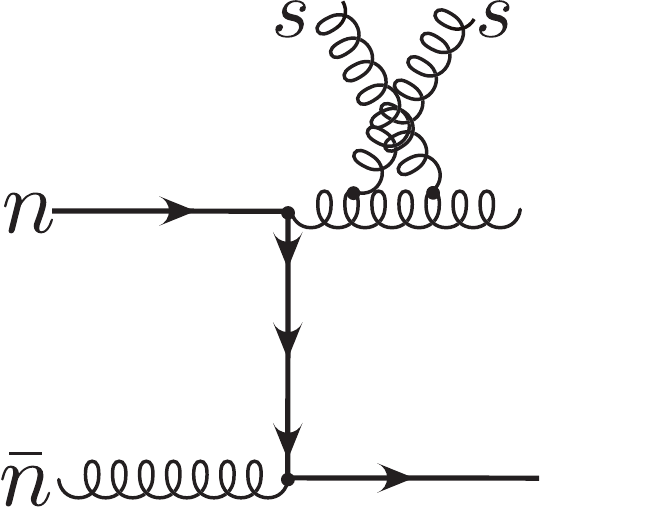}
				\hspace{0.2cm}
		\includegraphics[width=0.15\columnwidth]{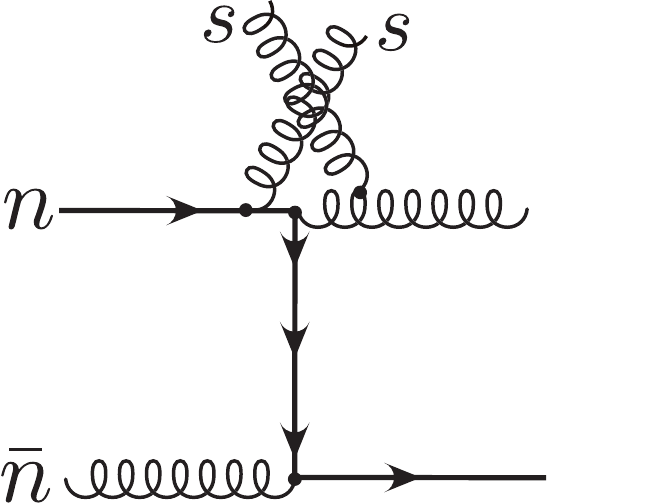}
				\hspace{0.2cm}
		\includegraphics[width=0.15\columnwidth]{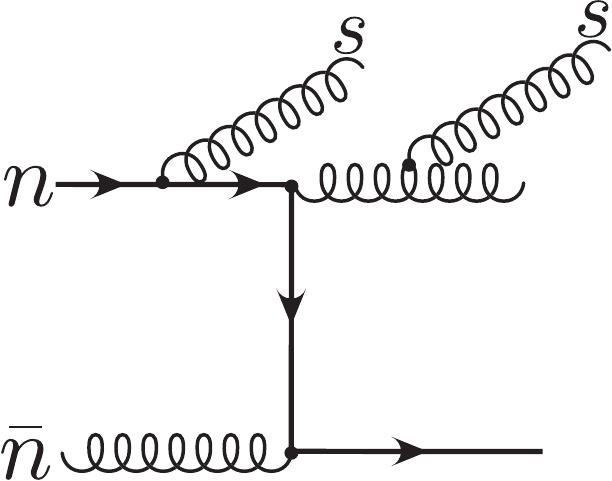}							
				\\[8pt]
		\hspace{-0.85cm}
		\includegraphics[width=0.15\columnwidth]{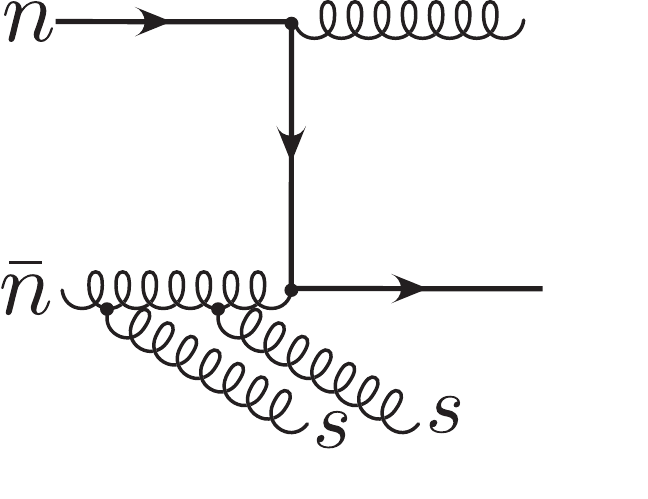}
		\hspace{0.2cm}
		\includegraphics[width=0.15\columnwidth]{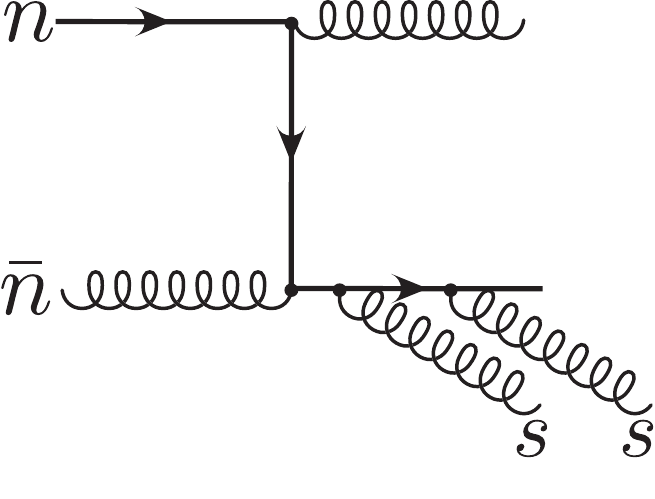}
		\hspace{0.2cm}
		\includegraphics[width=0.15\columnwidth]{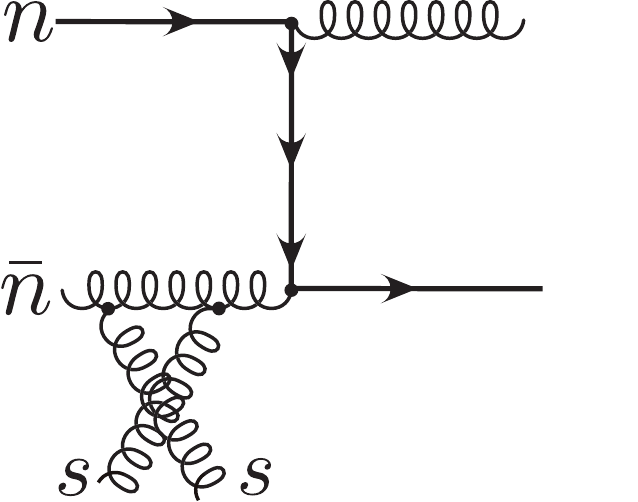}				
		\hspace{0.2cm}
		\includegraphics[width=0.15\columnwidth]{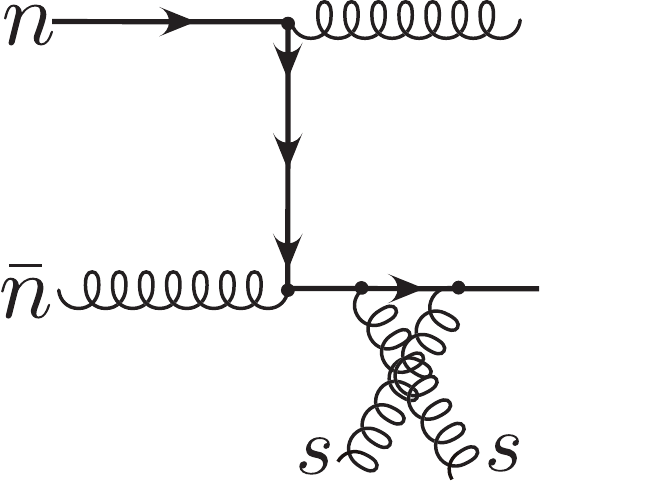}
				\hspace{0.2cm}
		\includegraphics[width=0.15\columnwidth]{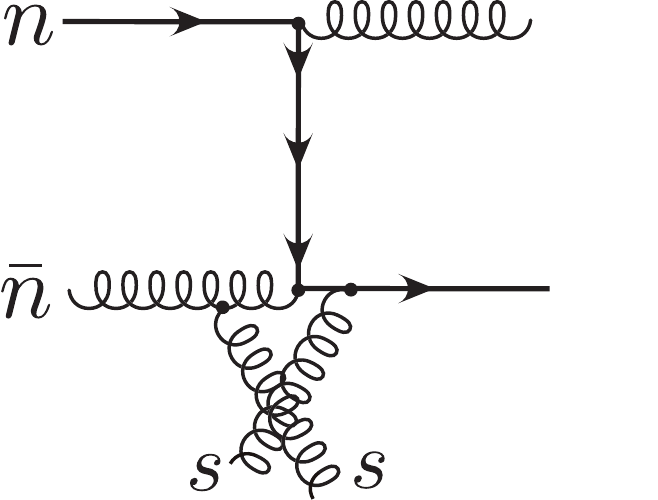}
				\hspace{0.2cm}
		\includegraphics[width=0.15\columnwidth]{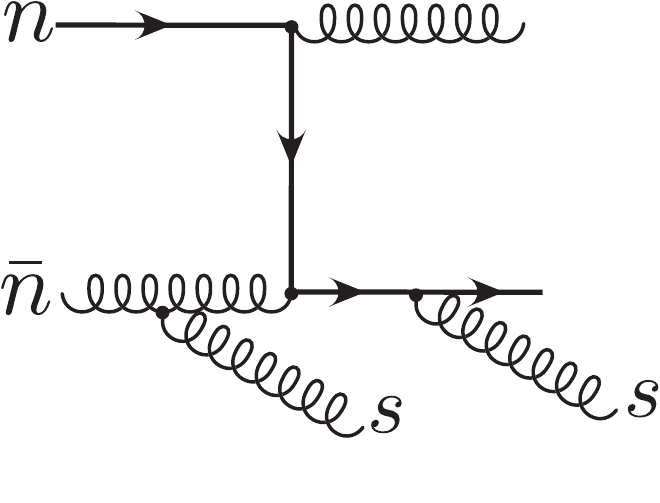}							
				\\[4pt]
		\hspace{-0.85cm}
		\raisebox{0.0cm}{\includegraphics[width=0.15\columnwidth]{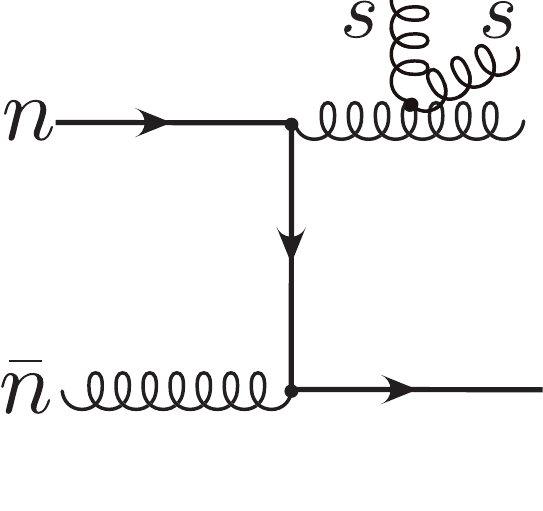}}
		\hspace{0.2cm}
		\raisebox{0.0cm}{\includegraphics[width=0.15\columnwidth]{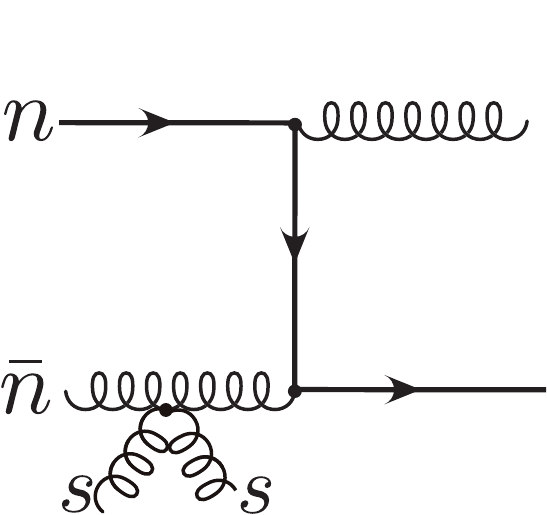}}			
		\hspace{0.2cm}
		\raisebox{0.5cm}{\includegraphics[width=0.15\columnwidth]{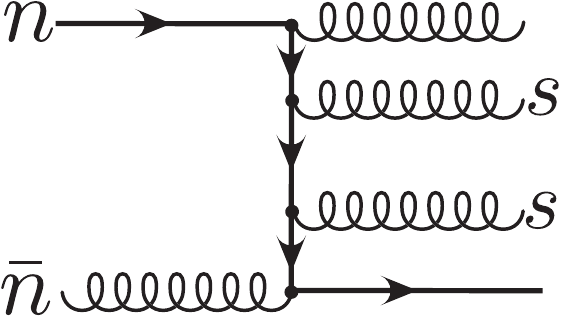}}
				\hspace{0.2cm}
		\raisebox{0.5cm}{\includegraphics[width=0.15\columnwidth]{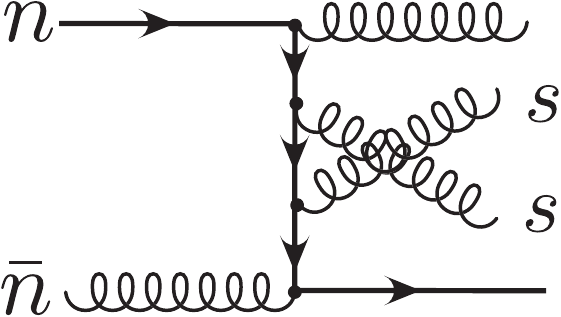}}						
		\\[8pt]
				\raisebox{2cm}{
			\hspace{-14.5cm}
			b)
		} \\[-50pt]	
		\hspace{-0.85cm}
		\includegraphics[width=0.18\columnwidth]{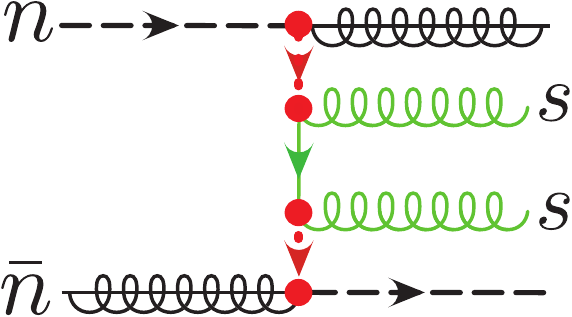}
		\hspace{0.2cm}
		\includegraphics[width=0.18\columnwidth]{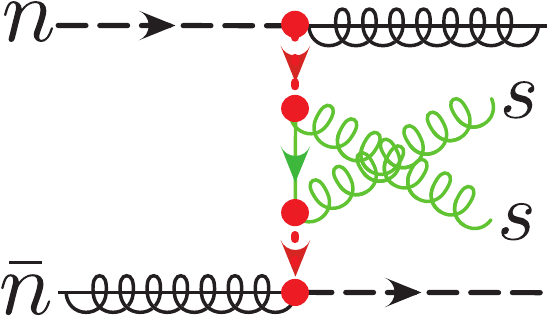}
		\hspace{0.2cm}
		\includegraphics[width=0.18\columnwidth]{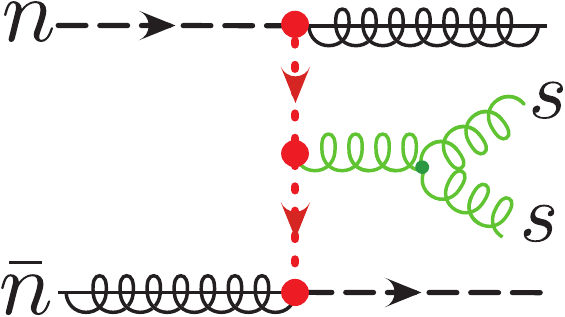}				
		\hspace{0.2cm}
		\includegraphics[width=0.18\columnwidth]{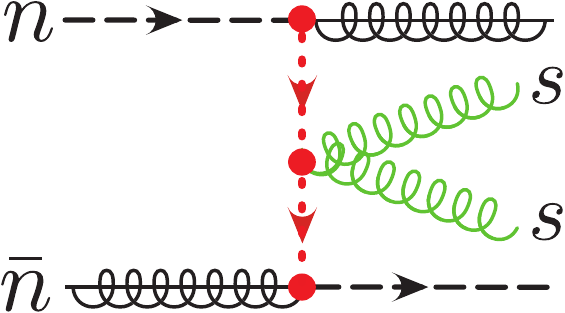}					
	\end{center}
	\vspace{-0.4cm}
	\caption{\setcaptionskip
		(a) Full theory and (b) effective theory graphs with two soft emissions. In the effective theory, the first three graphs are $T$-product contributions, and the fourth graph is the two emission Feynman rule from the Fadin-Sherman vertex.}
	\label{fig:double_emission}
	\setmainskip
\end{figure}

The double emission diagrams in the full theory and effective theory are shown in \Fig{fig:double_emission}. Note that the operators for $n$-$s$ and $\bar n$-$s$ forward scattering enter the matching through $T$-product contributions. 

Instead of performing the complete two emission matching, we will assume that only one ordering of Wilson lines appears, as in the case of the leading power Glauber Lagragian $\cL_G^{\text{II}(0)}$. This is motivated also by the patterns found in one emission matching as well as the structure of diagrams in \Fig{fig:double_emission} for the two emission matching. We leave a general proof of this statement to future work.
Under this assumption, we have $C_1=0$, which completely fixes the form of our soft operator to the final form given in \Eq{eq:soft_operator_quark}:
\begin{equation}
  \boxed{\cO_s = -2\pi \alpha_s \left[ S_{\bar n}^\dagger S_{n} \Sl{\cP}_\perp + \Sl{\cP}_\perp S_{\bar n}^\dagger S_{n} - S_{\bar n}^\dagger S_n g\Sl{\cB}^{n}_{S\perp}-  g\Sl{\cB}^{\bar n}_{S\perp} S_{\bar n}^\dagger S_n \right]}\,.
\end{equation}

The particular ordering of the Wilson lines, $S_\bn^\dagger S_n$, appearing in $\cO_s$ in \Eq{eq:soft_operator_quark} corresponds to the ordering of the collinear and soft operators in \Eq{eq:Glauber_Lagrangian}, and to the scattering configuration employed in our matching. The soft operator written with the opposite ordering is obtained simply by the replacement $n \leftrightarrow \bn$ in \Eq{eq:soft_operator_quark}.

\section{Quark Reggeization from Rapidity Renormalization}\label{sec:quark_reggeize}
In this section we consider the renormalization of the Glauber operators to derive the Reggeization of the quark. The renormalization should be done at the level of the squared amplitude, including both virtual and real contributions, to obtain IR finite results. Nevertheless, with careful interpretation of the IR divergences, the virtual diagrams can be examined at the amplitude level, and we will see that the solution to the rapidity renormalization group equation (RGE) corresponds to the Reggeization of the quark.

For quark-gluon scattering, we can decompose the color structure of the $t$-channel exchange as $3\, \otimes \, 8 = 3 \oplus  \bar 6 \oplus 15$.
Explicitly, if we decompose the amplitude using the color basis
\begin{align}\label{eq:Colordecomp}
\cM=2\left( T^A T^{B}\right)_{ij} \cA +2 \left( T^{B} T^A \right)_{ij} \cB +\delta^{AB} \delta_{ij} \cC  \,,
\end{align}
then the contributions to the $3$, $\bar 6$ and $15$ color structures are given by \cite{Bogdan:2002sr},
\begin{align}
\cM_3&=2C_F \cA -\frac{1}{N} \cB +\cC\,, \\
\cM_{\bar 6}&=-\cB+\cC\,, \\
\cM_{15}&=\cB+\cC \,.\label{eq:Colordecomp2}
\end{align}
In this section we will focus on the Reggeization of the $3$ channel at LL order, which corresponds to dressing the tree-level $t$-channel quark exchange. In the study of Reggeization, it is conventional to also decompose the amplitude so that it has a definite signature under crossing, i.e.,  $\cM^{\pm}=\frac{1}{2}\left[ \cM \pm \cM(s\leftrightarrow t)   \right]$. Indeed, it is known that it is the positive signature $3$ channel that builds upon the lowest order quark exchange and Reggeizes at LL order. The negative signature channel is suppressed by an $\alpha_s$, and has a series that starts at next-to-leading logarithmic (NLL) order, which is beyond the order we are working.

In \Sec{sec:consistency} we setup the notation and present the structure for the renormalization of the Glauber quark operators. We also derive consistency relations among the anomalous dimensions of the soft and collinear operators, which provide important checks on our calculation. In Secs.~\ref{sec:one_loop_collinear} and \ref{sec:one_loop_soft} we compute the anomalous dimension of the collinear and soft operators. In \Sec{sec:RG_solve} we solve the RGE and demonstrate the Reggeization of the quark.

The $\bar 6$ and $15$ channels are generated by the simultaneous exchange of both a Glauber quark and a Glauber gluon. These diagrams are rapidity finite at lowest order, and will be considered in \Sec{sec:glauber_box}.

\subsection{RG Structure and Consistency Relations}\label{sec:consistency}

For the collinear sector, there is no mixing and the renormalization has the structure
\begin{align}
\cO_n^{\, \bare} =V_{\cO_n} \cO_n \,, \qquad V_{\cO_n} =(1+\delta V_n) 
\,, 
\end{align}
with analogous relations for the $\bar{n}$ sector. Following~\cite{Rothstein:2016bsq}, we use the notation ``$V$'' instead of the traditional ``$Z$'' for renormalization factors to remind the reader that these are only virtual contributions and may still depend on IR regulator.

For the soft operator $\cO_s^n$, there is no mixing and we have
\begin{align}
\cO_{s}^{n \, \bare} = V_{\cO_s^n} \cO_s^n \,, \qquad V_{\cO_s^n}  =(1+\delta V_s^n) 
\,, 
\end{align}
with analogous relations for the $\bar{n}$ sector. For the soft operator $\cO_s$, the renormalization group structure is more complicated due to mixing with $T$-products of $\cO_s^n$ and $\cO_s^\bn$. 
This is discussed in detail for the Glauber gluon case in~\cite{Rothstein:2016bsq}.  The structure in our case is given by
\begin{align}
&\vec \cO_{s}^{ \, \bare}=\hat V_{\cO_{s}} \cdot \vec \cO_{s}\,, \nonumber \\
&\vec \cO_{s} =\left(\begin{array}{c} \cO_{s} \\ i\int d^4x~ T~ \cO_s^\bn(x) \bar{\cO}_s^n(0) \end{array} \right)\,, \quad 
\hat V_{\cO_{s}}=\left(\begin{array}{cc} 1+\delta V_s&0 \\ \delta V_s^T& 
\ V_{\cO_s^\bn}  V_{\bar{\cO}_s^n}  \end{array} \right)\,.
\end{align}
Importantly, due to the relative difference in the power counting of $\cO_s$ to that of $\cO_s^\bn$ and $\cO_s^n$, both components in $\vec \cO_{s}$ are the same order in the power counting. 

The renormalization group structure above, for both the collinear and soft sectors, is simpler than for the case of Glauber gluon operators, which involves mixing between quark and gluon operators that leads to the universality of Reggeization~\cite{Rothstein:2016bsq}. In the present case, there is only a non-trivial mixing in the soft sector.

The $\mu$ and $\nu$ anomalous dimensions are derived by demanding the $\mu$ and $\nu$ invariance of the bare operators as usual. Since our operators do not have Wilson coefficients and the soft and collinear fields are at the same $\mu$ scale, we expect their $\mu$ anomalous dimension to vanish, as in the case of $\cL_G^{(0)}$~\cite{Rothstein:2016bsq}. Therefore, we focus here on the $\nu$ anomalous dimensions, which give rise to rapidity renormalization, and the Reggeization. 

We have the standard relations
\begin{align}
\cO^{ \bare}= V_{\cO} \cdot \cO(\nu, \mu) \,, \qquad \nu \frac{\partial }{\partial \nu} \cO(\nu, \mu)=\gamma^\nu_{\cO} \cdot \cO(\nu, \mu) \,, \qquad \gamma^\nu_{\cO}=-V_{\cO}^{-1} \cdot \nu  \frac{\partial }{\partial \nu} V_{\cO} \,,
\end{align}
for $\cO = \cO_n \,, \cO_s^n\,, \cO_s$ and for the operators describing the $\bar{n}$ sector. 
For the soft operator $\cO_s$, which undergoes mixing, the anomalous dimension has the form
\begin{align}
\hat \gamma^\nu_{\cO_{s}}=\left(\begin{array}{cccc} \gamma_{s\nu}^{\text{dir}}&&&0 \\[2pt] 
\gamma_{s\nu}^{T}&&& \gamma^\nu_{\cO_s^\bn}  \gamma^\nu_{\bar{\cO}_s^n} \end{array} \right)\,.
\end{align}

The fact that there is no overall $\nu$ dependence in $n$-$\bar n$ scattering and $n$-$s$ scattering leads to relations among the anomalous dimensions. The consistency for $n$-$\bar n$ scattering is derived at the level of the time evolution operator, and one must consider all possible contributions from $T$-products involving $\cL_G^{\text{II}(0)}$, $\cL^{\text{II}(1/2)}$, and $\cL^{\text{II}(1)}$. At one-loop, this simplifies considerably, and we have
\begin{align}
\nu \frac{\partial }{\partial \nu}  \left( \cO_{\bn n} + i \int d^4x \ T \ \cO_{\bn s}(x)  \cdot \bar{\cO}_{n s}(0)  \right)=0\,.
\end{align}
Note again that this has homogeneous power counting. By differentiating the time evolution of the $n$-$\bar n$ scattering and the $n$-$s$ scattering, we can derive the following relations between anomalous dimensions
\begin{align}\label{eq:ADrelations}
\gamma^\nu_{\cO_n}=\gamma^\nu_{\cO_{\bar n}}\,, \qquad \gamma^{\text{dir}}_{s\nu} + \gamma^{T}_{s\nu}=-\gamma^\nu_{\cO_n}-\gamma^\nu_{\cO_{\bar n}}
\,, \qquad \gamma^\nu_{\cO_s^n}=-\gamma^\nu_{\cO_n}\,.
\end{align}

\subsection{One-Loop Virtual Anomalous Dimension for the Collinear Operator}\label{sec:one_loop_collinear}

\begin{figure}
\begin{center}
\begin{tabular}{cccc}
\fd{3cm}{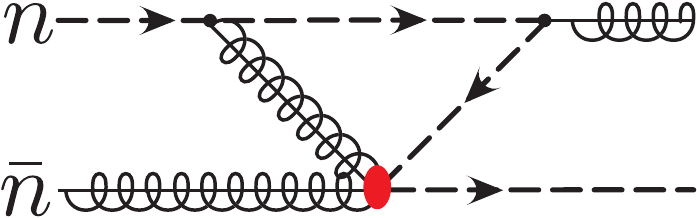} &
\fd{3cm}{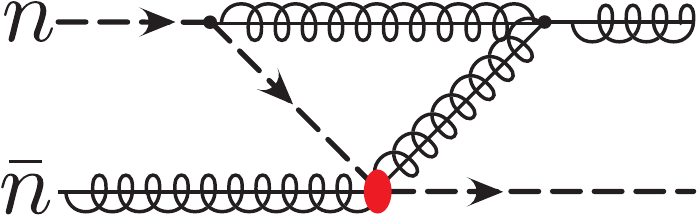} &
\fd{2cm}{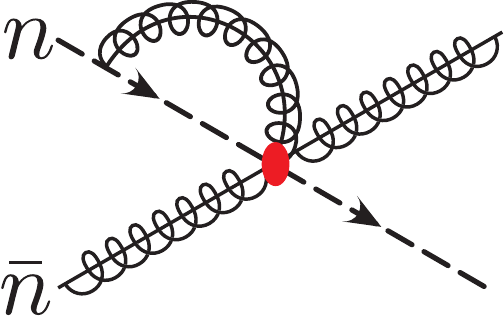} &
\fd{2cm}{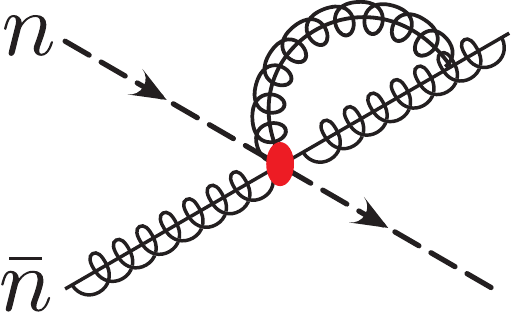} \\
a) & b) & c) & d)
\end{tabular}
\end{center}
\caption{One-loop virtual contributions to the renormalization of the collinear operator $\cO_n$. The V graphs are labeled a) and b), and the Wilson line graphs are labeled c) and d). 
}\label{fig:collinearAD}
\end{figure}

In this section we compute the one-loop virtual contributions to the renormalization of the collinear operator $\cO_n$. The two types of contributions are shown in Fig.~\ref{fig:collinearAD}, which we refer to as V graphs and Wilson line graphs. All the integrals can be evaluated following~\cite{Rothstein:2016bsq}, and we therefore only give the final results. It is sufficient to consider external gluons with perpendicular polarization, which  simplifies the calculation. We employ a gluon mass, $m$, as an IR regulator to ensure that all poles in $\epsilon$ are of UV origin. The IR regulator will explicitly appear in the rapidity anomalous dimension $\gamma^\nu_{\cO_n}$, and in the Regge trajectory.

In the following, we display only contributions to the $1/\eta$ pole (e.g., ignoring coupling and wavefunction renormalization), and denote finite pieces with ellipses. For the V graphs, we find
\begin{align}\label{eq:Vgraphs}
\text{Fig.~\ref{fig:collinearAD}a} &= (4 \pi \alpha_s)^2 (2C_F - C_A)  {\bar u}_{\bar n} \gamma_\perp^\mu T^A { \slashed{q}_\perp \over q_\perp^2} 
\int \dbar^d k  {\iota^\epsilon \mu^{2\epsilon}  |\bar n \cdot k|^{-\eta} \nu^\eta \slashed{k}_\perp (\slashed{k}_\perp + \slashed{q}_\perp) {\bar n} \cdot p_1   \over (k^2-m^2) (k+q)^2 (k+p_1)^2 {\bar n} \cdot k}  \gamma_\perp^\nu T^B u_n + \dots \nn  \\ 
&= -i 4 \pi \alpha_s {\bar u}_{\bar n} \gamma_\perp^\mu T^A { \slashed{q}_\perp \over q_\perp^2} \gamma_\perp^\nu T^B u_n \ {\alpha_s \over 2 \pi} \left( C_F - {C_A \over 2} \right)  {g (\epsilon, \mu^2/t) \over \eta} + \dots \,,
\\
\text{Fig.~\ref{fig:collinearAD}b} &= - (4 \pi \alpha_s)^2 C_A  {\bar u}_{\bar n} \gamma_\perp^\mu T^A { \slashed{q}_\perp \over q_\perp^2} 
 \int \dbar^d k   { \iota^\epsilon \mu^{2\epsilon} |\bar n \cdot k|^{-\eta} \nu^\eta \slashed{k}_\perp (\slashed{k}_\perp + \slashed{q}_\perp) {\bar n} \cdot p_4   \over (k^2-m^2) (k+q)^2 [( k - p_4)^2 - m^2] {\bar n} \cdot k} \gamma_\perp^\nu T^B u_n  + \dots \nn \\ 
&= -i 4 \pi \alpha_s {\bar u}_{\bar n} \gamma_\perp^\mu T^A { \slashed{q}_\perp \over q_\perp^2} \gamma_\perp^\nu T^B u_n   \ {\alpha_s \over 2 \pi} {C_A \over 2}   {g (\epsilon, \mu^2/t) \over \eta} + \dots\,,
\end{align}
where
\begin{align}
g(\epsilon, \mu^2/t) = e^{\epsilon \gamma_E} \left(\mu^2 \over -t \right)^\epsilon \cos(\pi \epsilon) \Gamma(-\epsilon) \Gamma(1+2 \epsilon) \,.
\end{align}
These results are independent of the IR regulator $m$, with $t$ regulating the IR region. For the Wilson line graphs, we find
\begin{align}
\text{Fig.~\ref{fig:collinearAD}c} &= - (4 \pi \alpha_s)^2 (2C_F - C_A)  \int \dbar^d k  { \iota^\epsilon \mu^{2\epsilon} |\bar n \cdot k|^{-\eta} \nu^\eta  {\bar n} \cdot p_1   \over (k^2 - m^2)  (k+p_1)^2 {\bar n} \cdot k}     {\bar u}_{\bar n} \gamma_\perp^\mu T^A { \slashed{q}_\perp \over q_\perp^2} 
\gamma_\perp^\nu T^B u_n + \dots \nn  \\ 
&= -i 4 \pi \alpha_s {\bar u}_{\bar n} \gamma_\perp^\mu T^A { \slashed{q}_\perp \over q_\perp^2} \gamma_\perp^\nu T^B u_n \ {\alpha_s \over 2 \pi} \left( C_F - {C_A \over 2} \right)  {h (\epsilon, \mu^2/m^2) \over \eta}  + \dots \,, \\
\text{Fig.~\ref{fig:collinearAD}d} &= - (4 \pi \alpha_s)^2  C_A  \int \dbar^d k  {\iota^\epsilon \mu^{2\epsilon} |\bar n \cdot k|^{-\eta} \nu^\eta  {\bar n} \cdot p_4   \over (k^2 - m^2)  (k+p_4)^2 {\bar n} \cdot k}     {\bar u}_{\bar n} \gamma_\perp^\mu T^A { \slashed{q}_\perp \over q_\perp^2} 
\gamma_\perp^\nu T^B u_n  +\dots \nn  \\ 
&= -i 4 \pi \alpha_s {\bar u}_{\bar n} \gamma_\perp^\mu T^A { \slashed{q}_\perp \over q_\perp^2} \gamma_\perp^\nu T^B u_n \ {\alpha_s \over 2 \pi} {C_A \over 2}  {h (\epsilon, \mu^2/m^2) \over \eta}  + \dots \,,
\end{align}
where
\begin{align}
h(\epsilon, \mu^2/m^2) = e^{\epsilon \gamma_E} \left(\mu^2 \over m^2\right)^\epsilon  \Gamma(\epsilon)  \,.
\end{align}
Here we see an explicit dependence on the IR regulator $m$.
Note that the $C_A$  dependence of the $1/\eta$ pole cancels in the sum for both the V graphs and Wilson line graphs. Upon summing all diagrams in Fig.~\ref{fig:collinearAD}, we find
\begin{align}
\delta V_n &=  { \alpha_s C_F  \over 2\pi} \left[ {g (\epsilon, \mu^2/t)+ h (\epsilon, \mu^2/m^2) \over \eta} \right] \,,\nn \\
\gamma^\nu_{\cO_n} &= { \alpha_s C_F  \over 2\pi} \bigg[ g (\epsilon, \mu^2/t)+ h (\epsilon, \mu^2/m^2) \bigg] = { \alpha_s C_F  \over 2\pi} \text{ln}\left( -t \over m^2 \right) \label{eq:gamman}\,,
\end{align}
where we expanded in $\epsilon$ in the final result for $\gamma^\nu_{\cO_n} $. This result is the same as for the Glauber gluon case up to Casimir scaling.

\subsection{One-Loop Virtual Anomalous Dimension for the Soft Operator}\label{sec:one_loop_soft}
\begin{figure}
\begin{center}
\begin{tabular}{cc}
\fd{3.4cm}{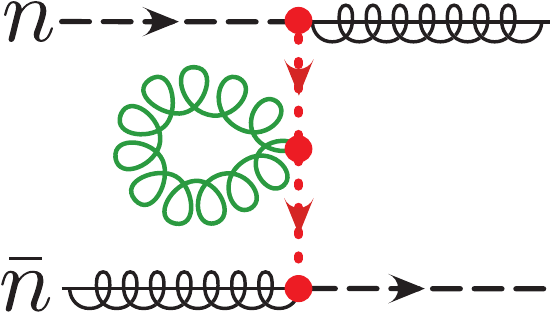} &
\fd{2.6cm}{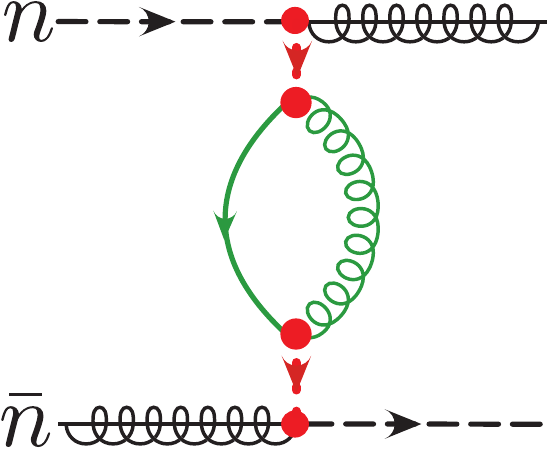} \\[30pt]
a) & b)
\end{tabular}
\vspace{-0.4cm}
\end{center}
\caption{One-loop virtual contributions to the renormalization of the soft operator $\cO_s$. The flower graph is labeled a) and the eye graph is labeled b).
}\label{fig:softAD}
\end{figure}

The result for the anomalous dimension $\gamma^\nu_{\cO_n}$ in Eq.~(\ref{eq:gamman}), along with the relations in Eq.~(\ref{eq:ADrelations}), specify the complete set of anomalous dimensions for our operators. Nonetheless, in this section we explicitly compute the renormalization of the soft operator $\cO_s$, verifying the structure of the operator mixing and the result for the combination $\gamma^{\text{dir}}_{s\nu} + \gamma^{T}_{s\nu}$. 

The relevant diagrams are shown in \Fig{fig:softAD}, which we refer to as the flower graph, and the eye graph. As in the previous section, all integrals can be performed using techniques from~\cite{Rothstein:2016bsq}, so we present only the final results, and again we keep only terms that contribute to the $1/\eta$ pole, as required for the rapidity renormalization. 
For the flower diagram, we find 
\begin{align}
\text{Fig.~\ref{fig:softAD}a}&= - (4 \pi \alpha_s)^2 2 C_F {\bar u}_{\bar n} \gamma_\perp^\mu T^A { \slashed{q}_\perp \over q_\perp^2} \gamma_\perp^\nu T^B u_n  \int \dbar^d k \frac{\iota^\epsilon \mu^{2\epsilon} |2k_z|^{-\eta} \nu^\eta}{(k^2-m^2) n\cdot k \bar n \cdot k}  +\dots \nn \\
&=  -i 4 \pi \alpha_s {\bar u}_{\bar n} \gamma_\perp^\mu T^A { \slashed{q}_\perp \over q_\perp^2} \gamma_\perp^\nu T^B u_n \left[  - {\alpha_s \over \pi} C_F  {h (\epsilon, \mu^2/m^2) \over \eta}   \right]+\dots \,.
\end{align}
For the eye diagram, we find
\begin{align}
\text{Fig.~\ref{fig:softAD}b}&=  -(4\pi\alpha_s)^2 2 C_F  {\bar u}_{\bar n} \gamma_\perp^\mu T^A { \slashed{q}_\perp \over q_\perp^2} \left[ \int \dbar^d k {\iota^\epsilon \mu^{2\epsilon} |2k_z|^{-\eta} \nu^\eta  \slashed{k}_\perp (\slashed{k}_\perp + \slashed{q}_\perp) \slashed{k}_\perp \over  (k^2-m^2) (k + q)^2 n \cdot k {\bar n} \cdot k } \right]  { \slashed{q}_\perp \over q_\perp^2} \gamma_\perp^\nu T^B u_n +\dots \nn \\
&= -i 4 \pi \alpha_s {\bar u}_{\bar n} \gamma_\perp^\mu T^A { \slashed{q}_\perp \over q_\perp^2} \gamma_\perp^\nu T^B u_n  \left[  - {\alpha_s \over \pi} C_F  {g (\epsilon, \mu^2/t) \over \eta}   \right]  +\dots \,. 
\end{align}
These results determine the counterterms and anomalous dimensions as
\begin{align}
\delta V_s &= - {\alpha_s \over \pi} C_F  {h (\epsilon, \mu^2/m^2) \over \eta} \,, \qquad 
\delta V_s^T = - {\alpha_s \over \pi} C_F  {g (\epsilon, \mu^2/t) \over \eta} \,, \nn\\
\gamma_{s \nu}^\text{dir} &=  - {\alpha_s \over \pi} C_F  h (\epsilon, \mu^2/m^2)\,, \quad \gamma_{s \nu}^T = - {\alpha_s \over \pi} C_F  g (\epsilon, \mu^2/t)  \, ,
\end{align}
consistent with those for the collinear sector. In the next section, we will solve the RGE and see that the anomalous dimension fixes the form of the Regge trajectory.

\subsection{Solving the Rapidity RGE}\label{sec:RG_solve}

With the anomalous dimensions in hand, it is now straightforward to achieve amplitude level Reggeization through solving the rapidity RGE.
We have the rapidity anomalous dimensions $\gamma^\nu_{\cO_n}$ for the collinear operator $\cO_n$ and $\gamma^{\text{dir}}_{s\nu} + \gamma^{T}_{s\nu}$ for the soft operator $\cO_s$, which satisfy the required consistency relations in Eq.~(\ref{eq:ADrelations}), This ensures that we can equivalently either run the collinear operators to the soft scale, or the soft operators to the collinear scale. We choose to run the collinear operators to the soft scale. The rapidity RGE is given by
\begin{align}
\nu \frac{d}{d\nu} \cO_n(\nu) =\gamma^\nu_{\cO_n} \cO_n(\nu)\,,
\end{align}
where the argument explicitly denotes the dependence on the $\nu$ scale (the $\mu$ scale does not enter our analysis). Since the anomalous dimension is independent of $\nu$, the solution is
\begin{align}
\cO_n\big(\sqrt{-t}\big)=\left( \frac{s}{-t}  \right)^{-\frac12 \gamma^\nu_{\cO_n} } \cO_n\big(\sqrt{s}\big)\,,
\end{align}
with an analogous expression for the $\bn$-collinear sector. Upon substituting the evolved collinear operators into the forward scattering operator, we find
\begin{align}\label{eq:evolvedscattering}
\cO_{n\nbar}\big(\sqrt{-t}\big)
&=\left( \frac{s}{-t}  \right)^{-\frac{\alpha_s(\mu)C_F}{2\pi}  \log\left( \frac{-t}{m^2}  \right)}    \bar{\cO}_n\big(\sqrt{s}\big) \frac{1}{\Sl{\cP}_\perp}  \cO_s\big(\sqrt{-t}\big)   \frac{1}{\Sl{\cP}_\perp} \cO_{\bar n}\big(\sqrt{s}\big) \,,
\end{align}
which is the one-loop Reggeization of the quark. We emphasize again that we have not decomposed this result into amplitudes of definite signature. At LL order, $\log(s/|t|)$ and $\log(-s/|t|)$ are equivalent, and only differ at NLL order. The one-loop Regge trajectory for the quark is given by the exponent in \Eq{eq:evolvedscattering}:
\begin{equation}
\omega_q=-\frac{\alpha_s(\mu)C_F}{2\pi}  \log\left( \frac{-t}{m^2}  \right)\,,
\end{equation}
which agrees with the known result~\cite{Fadin:1976nw,DelDuca:2011ae}. Here it emerges directly from the rapidity renormalization of operators in the SCET subleading power Lagrangian. The one-loop quark Regge trajectory is identical to that for the gluon up to Casimir scaling, $C_A \to  C_F$. In a physical cross section, the dependence on the IR cutoff $m$ is cancelled by real emission diagrams, leading to an IR finite result. In \Sec{sec:quark_BFKL}, we will consider Reggeization at the cross section level for $q\bar q \to \gamma \gamma$, which will lead to the IR finite BFKL equation.

\subsection{Glauber Boxes}\label{sec:glauber_box}

\begin{figure}
\begin{center}
\begin{tabular}{cccc}
\hspace{-0.4cm} \fd{3.5cm}{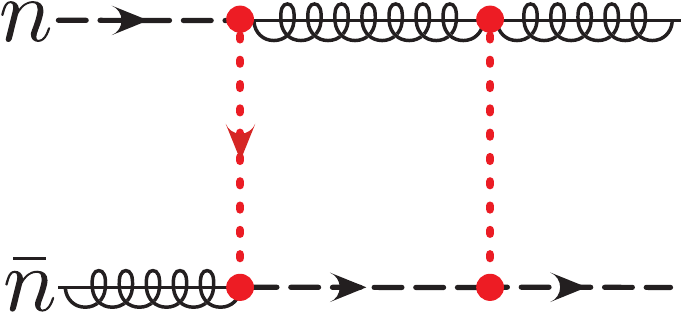}  &
\hspace{0.15cm}\fd{3.5cm}{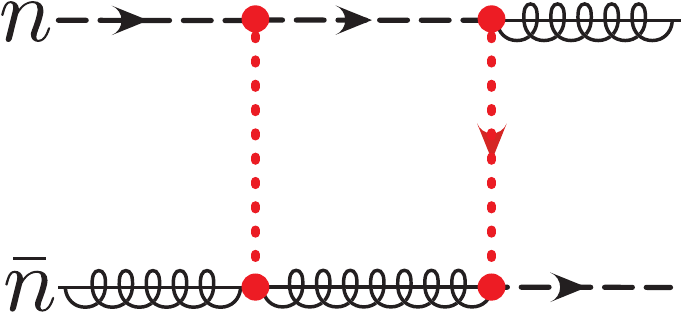} &
 \hspace{0.15cm}\fd{3.5cm}{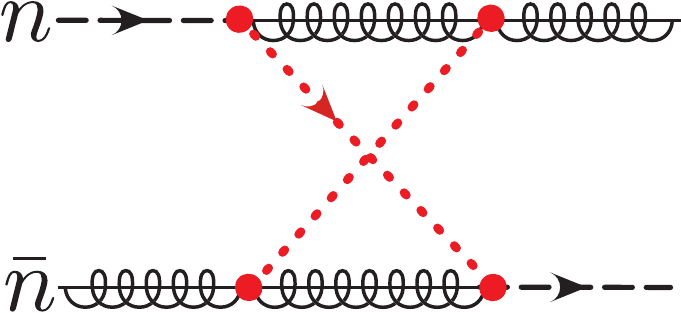} &
\hspace{0.15cm}\fd{3.5cm}{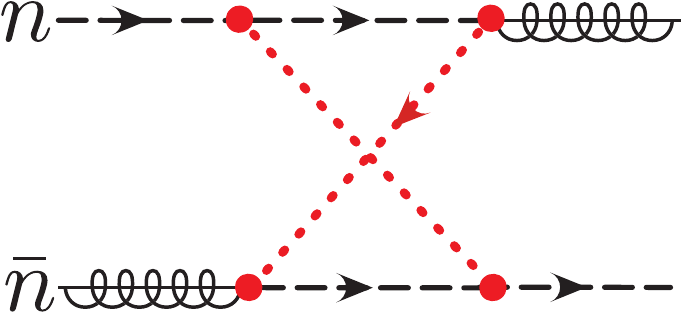}  
\\[25pt] 
a)  & b) &
  c) & d)
\end{tabular}
\vspace{-0.5cm}
\end{center}
\caption{Graphs contributing to the $\bar 6$ and $15$ color structures of the t-channel exchange. The cross box diagrams labeled c) and d) vanish with our regulator. 
}\label{fig:boxes}
\end{figure}

So far, in this section we have focused on the structure of the rapidity divergent contributions, which lead to the Reggeization of the $3$ color channel. At $\cO(\alpha_s^2)$ there are also non-vanishing contributions to the $\bar 6$ and $15$ color channels, which are known in the literature~\cite{Bogdan:2002sr}. In this section, we show that these are reproduced in a very simple manner in our framework by the simultaneous exchange of a Glauber quark and a Glauber gluon, as shown in \Fig{fig:boxes}.

As discussed in detail in~\cite{Rothstein:2016bsq}, the box graphs with Glauber scaling for the loop momentum require the rapidity regulator $|2k^z|^{-\eta} \nu^\eta$ to make them well defined (but are independent of $\eta$ as $\eta \to 0$). In particular, the Glauber cross box diagram vanishes due to having poles in $k^0$ on the same side of the contour. This is crucial since the box and cross box diagrams have different color factors, and thus illustrates the nontrivial mapping between the calculations in the EFT defined with our regulator, and full QCD. The ability to reproduce the known results for the $\bar 6$ and $15$ channels therefore provides a non-trivial test of the regulator, and of the EFT simultaneously involving quark and gluon Glauber operators.

Since the Glauber cross boxes shown in Fig.~\ref{fig:boxes}c and Fig.~\ref{fig:boxes}d vanish, we only compute the boxes shown in Fig.~\ref{fig:boxes}a and Fig.~\ref{fig:boxes}b. The $k^0$ and $k^z$ integrations are the same as for the box graphs with only Glauber gluons considered in~\cite{Rothstein:2016bsq}, while the $k_\perp$ integration is modified by the presence of the Glauber quark. Employing the results for the integrals in~\cite{Rothstein:2016bsq}, we find
\begin{align}
\text{Fig.~\ref{fig:boxes}a}   &= -\delta^{AB}\delta_{ij}   2 \pi^2 \alpha_s^2 \bar{u}_{\bar n}  \gamma_\perp^\mu  \left[ \left( -i \over 4\pi \right)\int  {\dbar^{d-2} k_\perp \slashed{k}_\perp (-i \pi) \over \vec{k}_\perp^2  (\vec{k}_\perp +\vec{q}_\perp)^2}    \right] \gamma_\perp^\nu u_n  \,, \\
\text{Fig.~\ref{fig:boxes}b} &=    \delta^{AB} \delta_{ij}  2 \pi^2 \alpha_s^2 \bar{u}_{\bar n}  \gamma_\perp^\mu \left[ \left( -i \over 4\pi \right) \int {\dbar^{d-2} k_\perp(\slashed{k}_\perp + \slashed{q}_\perp) (-i \pi) \over \vec{k}_\perp^2  (\vec{k}_\perp +\vec{q}_\perp)^2}   \right] \gamma_\perp^\nu u_n \,.
\end{align}
Just like for the exchange of two Glauber gluons, these box diagrams yield ``$i\pi$" factors that are characteristic of Glauber loops. Here we have simplified the color structure as $(T^D T^A T^C)_{ij} f^{BCD} = i\delta^{AB} \delta_{ij}/4 $. The sum of the diagrams is
\begin{align}
\text{Fig.~\ref{fig:boxes}a}    + \text{Fig.~\ref{fig:boxes}b}   &=   \left[ -i 4 \pi \alpha_s \bar{u}_{\bar n}  \gamma_\perp^\mu { \slashed{q}_\perp \over q_\perp^2} \gamma_\perp^\nu u_n \right]   \delta^{AB} \delta_{ij} { \alpha_s \over 4\pi} \left[ -{1 \over \epsilon} - \log { \mu^2 \over -t} \right]   (-i \pi) \, . \label{eq:boxes}
\end{align}
From Eq.~(\ref{eq:boxes}) we find a nonzero contribution to the color amplitude $\cC$ in the decomposition of Eqs.~(\ref{eq:Colordecomp}-\ref{eq:Colordecomp2}), and thus the contributions to the $\bar 6$ and $15$ color structures are
\begin{align}
\cM_{\bar 6} = \cM_{15} =  \left[ -i 4 \pi \alpha_s \bar{u}_{\bar n}  \gamma_\perp^\mu { \slashed{q}_\perp \over q_\perp^2} \gamma_\perp^\nu u_n \right] { \alpha_s \over 4\pi} \left[ -{1 \over \epsilon} - \log { \mu^2 \over -t} \right]   (-i \pi) \, ,
\end{align}
which agrees with the results of~\cite{Bogdan:2002sr} upon accounting for conventions.

\section{BFKL for $q\bar q\to \gamma \gamma$}\label{sec:quark_BFKL}

In this section we consider the application of Glauber quark operators for $q\bar q\to \gamma \gamma$ forward scattering. In QED, fermion Reggeization in the process $e^+e^- \to \gamma \gamma$ was studied in~\cite{Sen:1982xv}. Here we will follow the framework laid out in~\cite{Rothstein:2016bsq}, where the BFKL equation was derived from the rapidity renormalization of Glauber gluon operators at the cross section level. With Glauber operators in the effective theory, one can no longer factorize soft and collinear dynamics to all orders. However, with any fixed number of Glauber exchanges, the factorization is still possible, and therefore one can consider an expansion in the number of Glauber operator insertions. The first term in this expansion has a single Glauber gluon on either side of the cut and is referred to as the Low-Nussinov Pomeron approximation. This was used in~\cite{Rothstein:2016bsq} to derive the BFKL equation at LL order.

Unlike for the gluon BFKL, where one must consider an arbitrary number of Glauber operator insertions, for the case of quark Reggeization, the Glauber quark operators have an explicit power suppression, and therefore cannot be iteratively inserted. Instead, we must consider a single quark Glauber operator insertion on either side of the cut plus an arbitrary number of Glauber gluon operator insertions with $\cL_G^{(0)}$. To proceed, one must therefore still expand in the number of leading power Glauber gluon exchanges. To LL accuracy the situation simplifies significantly, and we only need to consider the factorization of the forward scattering matrix element with a single quark Glauber insertion on either side of the cut. Following \cite{Rothstein:2016bsq}, we can write the transition matrix element as
\begin{align}
T^q_{(1,1)} = \int d^2 q_\perp d^2 q^\prime_\perp C^q_n(q_\perp, p^-) S^q(q_\perp, q_\perp^\prime) C^q_{\bar n} (q^\prime_\perp, p^{\prime +}) \,,
\end{align}
where $C^q_n(q_\perp, p^-)$ and $C^q_{\bar n} (q^\prime_\perp, p^{\prime +})$ are squared collinear matrix elements and $S^q(q_\perp, q_\perp^\prime)$ is a squared soft matrix element. The subscript $(1,1)$ indicates that there is a single quark Glauber exchange on either side of the cut and the $q$ superscript distinguishes these matrix elements from the matrix elements of operators of $\cL_G^{(0)}$ describing Glauber gluon exchange from \cite{Rothstein:2016bsq}. In evaluating the matrix elements above, large logs arise due to the interplay of collinear modes whose natural rapidity scale is $\sqrt{s}$ and soft modes whose natural rapidity scale is $\sqrt{-t}$. We will resum these logs by considering the renormalization of the transition amplitude $T^q_{(1,1)}$ at LL order, and we will find that the resulting evolution equation is the same as the BFKL equation~\cite{Kuraev:1977fs,Balitsky:1978ic} up to Casimir scaling.

\begin{figure}
\begin{center}
\begin{tabular}{ccc}
\fd{3.2cm}{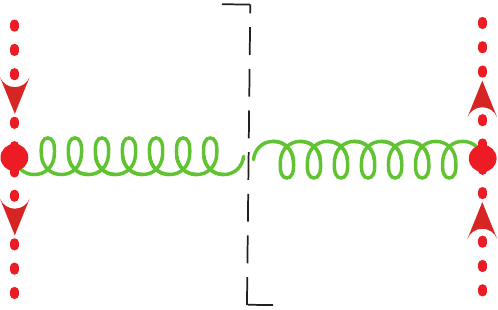}  &
\hspace{1.0cm}\fd{2.6cm}{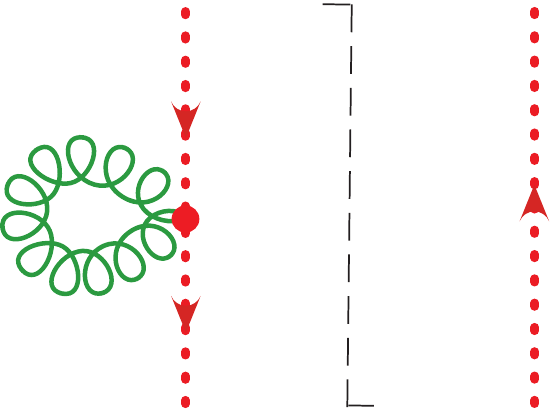} &
\hspace{1.0cm}\fd{2.6cm}{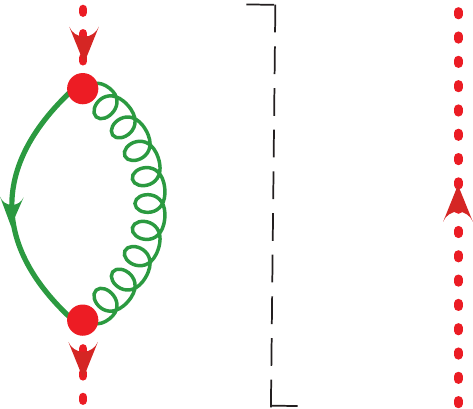} 
\\[30pt]
a)  & \hspace{1.3cm} b)  & \hspace{1.2cm} c)
\end{tabular}
\vspace{-0.5cm}
\end{center}
\caption{Graphs contributing to the LL order evolution of the soft function $S(q_\perp, q_\perp^\prime)$. The real contribution is labeled a), and the virtual contributions are labeled b) and c). The black dashed line represents the final state cut.
}\label{fig:BFKLgraphs}
\end{figure}

\subsection{BFKL Equation for the Soft Function}
Let us choose the rapidity scale in the renormalized transition matrix element  $T^q_{(1,1)}(\nu)$ to be $\nu = \sqrt{s}$, and consider the running of the soft function from $\nu= \sqrt{-t}$ to $\nu=\sqrt{s}$ to resum the large logs. This requires the one-loop real and virtual diagrams shown in Fig.~\ref{fig:BFKLgraphs}. In addition to these diagrams, there are also diagrams involving a Glauber gluon, and real soft quarks crossing the cut, coming from a power suppressed SCET Lagrangian. It is straightforward to show that such contributions are not rapidity divergent, which is expected, since the analogous virtual graphs are not associated with the Reggeization of the quark. For the calculations in this section we drop the mass regulator since IR divergences will cancel between the real and virtual contributions, and we set $d=4$ since only rapidity divergences are relevant for our analysis.

We define the soft function as
\begin{align}\label{eq:BFKLsoft}
S^q(q_\perp,q^\prime_\perp) = -{(2\pi)^4 \over V_2} {\delta^{ii'} \delta^{j j'} \over 
q^\mu_\perp q^{\prime \nu}_\perp \gamma^{\{ \mu}_{\alpha \bar{\alpha}} \gamma_{\beta \bar{\beta}}^{ \dagger \nu\} }} \sum_X \langle 0 | \cO_{s \alpha \bar{\alpha}} ^{ij}(q_\perp, q^\prime_\perp)  | X \rangle  \langle X | \cO_{s \beta \bar{\beta}}^{ \dagger i^\prime j^\prime}(q_\perp, q^\prime_\perp) |0 \rangle \,,
\end{align}
where the volume factor is $V_2 =(2\pi)^2 \delta^2(0) $, the color indices $i,j,i^\prime, j^\prime$ and fermionic indices $\alpha, {\bar \alpha}, \beta , {\bar \beta}$ have been made explicit, and for normalization we divide out by  $- q^\mu_\perp q^{\prime \nu}_\perp \gamma^{\{ \mu}_{\alpha \bar{\alpha}} \gamma_{\beta \bar{\beta}}^{ \dagger \nu\} }  = -\frac12 \{ \slashed{q}_\perp ^\prime \slashed{q}^\dagger_\perp + \slashed{q}_\perp \slashed{q}^{\prime \dagger}_\perp  \}$. 

We now compute the tree level and one-loop real and virtual contributions to the soft function. At tree level, the matrix element of the soft operator and the soft function obtained from squaring it are
\begin{align}\label{eq:BFKLtree}
\langle 0 | \cO_s^{ij}  | 0 \rangle = - i 4\pi \alpha_s \slashed{q}_\perp \delta^2(\vec{q}_\perp + \vec{q}^\prime_\perp ) \delta^{ij} \,, \qquad S^q_0(q_\perp,q^\prime_\perp) = (4\pi \alpha_s)^2 \delta^{ii} (2\pi)^2 \delta^2(\vec{q}_\perp + \vec{q}^\prime_\perp ) \, .
\end{align}

For the $\cO(\alpha_s)$ real contribution shown in Fig.~\ref{fig:BFKLgraphs}a, we compute the square of the one-gluon Feynman rule from the Fadin-Sherman vertex. Upon summing over gluon polarizations in Feynman gauge, we find
\begin{align}
{(2\pi)^4 \over V_2} \langle 0 | \cO_S^{ij}  | g \rangle  \langle g | \cO_S ^{ij \dagger}   | 0 \rangle = -(4\pi \alpha_s)^3 2C_F \delta^{ii} (2\pi)^2 \delta^2(\vec{q}_\perp + \vec{q}^\prime_\perp + \vec{k}_\perp ) { \{ \slashed{q}_\perp ^\prime \slashed{q}^\dagger_\perp + \slashed{q}_\perp \slashed{q}^{\prime \dagger}_\perp  \}  \over n \cdot k {\bar n} \cdot k}+\cdots\,,
\end{align}
where we have dropped the term having $\gamma^\mu_{\alpha \bar{\alpha}\perp} \gamma^{\dagger \mu}_{\beta \bar{\beta} \perp }$, which is rapidity finite. Using this result in Eq.~(\ref{eq:BFKLsoft}) we find the contribution to the soft function
\begin{align}
S_{1}^{q,\text{real}} = {\alpha_s C_F \over \pi^2 } \Gamma \left[ \eta \over 2 \right] \int {d^2k_\perp \over (\vec{k}_\perp - \vec{q}_\perp)^2 } S^q_{0}(k_\perp, q^\prime_\perp) +\cdots \, ,
\end{align}
where we have included the integral over phase space and identified the tree-level soft function. The ellipses denote rapidity finite contributions that will not play a role in the rapidity renormalization.

For the virtual corrections, we have the same flower and eye graphs appearing in the analysis for quark Reggeization in Sec.~\ref{sec:one_loop_soft}. As before, we keep only rapidity divergent contributions. The flower graph, appearing in Fig.~\ref{fig:BFKLgraphs}b, is given by
\begin{align}
\fd{1.0cm}{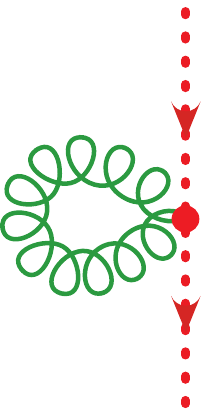} = -2(4 \pi \alpha_s)^2  C_F \delta_{ij} \int \dbar^4 k { w^2 |2 k_z|^{-\eta} \nu^{\eta} \slashed{q}_\perp \over k^2~ n \cdot k~ \bar{n} \cdot k}  \delta^2(\vec{q}_\perp + \vec{q}^\prime_\perp ) + \dots \, ,
\end{align}
where the ellipses denote rapidity finite terms. The eye graph, appearing in Fig.~\ref{fig:BFKLgraphs}c, is given by
\begin{align}
\fd{0.85cm}{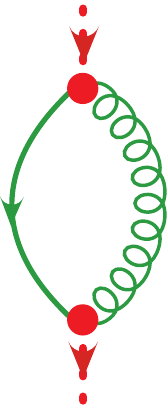} &=- 2(4\pi \alpha_s)^2  C_F \delta_{ij} \int \dbar^4 k { w^2 |2 k_z|^{-\eta} \nu^{\eta} \slashed{k}_\perp (\slashed{k} + \slashed{q}_\perp) \slashed{k}_\perp \over k^2~ (k+q_\perp)^2 ~n \cdot k~ \bar{n} \cdot k} \delta^2(\vec{q}_\perp + \vec{q}^\prime_\perp ) +\cdots  \\
&=2 (4\pi \alpha_s)^2 C_F \delta_{ij}  \int \dbar^4 k { w^2 |2 k_z|^{-\eta} \nu^{\eta} \over n \cdot k~ \bar{n} \cdot k} \left[  {\slashed{q}_\perp \over (k+q_\perp)^2}   + {q_\perp^2 \slashed{k} \over k^2 (k+q_\perp)^2} \right]  \delta^2(\vec{q}_\perp + \vec{q}^\prime_\perp ) +\cdots \,, \nn
\end{align}
where in the second line we dropped integrands that are odd in $k$. Note that the first term in the square brackets cancels the flower graph. The total virtual contribution is then
\begin{align}
\fd{1cm}{figures_BFKL/fermionic_flower_BFKL_oneside_low.pdf} \ \ +  \ \fd{0.85cm}{figures_BFKL/fermionic_eye_BFKL_oneside_low.pdf} \ = i 4\pi \alpha_s^2 C_F \delta_{ij} \Gamma \left[ \eta \over 2 \right] \int \dbar^2 k_\perp {\vec{q}_\perp^{\,\,2} \slashed{q}_\perp  \over \vec{k}_\perp^{\,2} (\vec{k}_\perp - \vec{q}_\perp)^2  }  \delta^2(\vec{q}_\perp + \vec{q}^\prime_\perp ) +\cdots\, .
\end{align}
We combine this result with the tree-level matrix element in Eq.~(\ref{eq:BFKLtree}) to obtain the squared matrix element. Hence the one-loop virtual contribution to the soft function is 
\begin{align}
S_{1}^{q,\text{virtual}}= -  {\alpha_s C_F \over 2 \pi^2}  \Gamma \left[ \eta \over 2 \right] \int d^2 k_\perp {\vec{q}_\perp^{\,\,2}  \over \vec{k}_\perp^{\,2} (\vec{k}_\perp - \vec{q}_\perp)^2  } S^q_0(q_\perp, q^\prime_\perp) +\cdots  \,. 
\end{align}

These results for the real and virtual corrections, $S_{1}^{q,\text{real}}$ and $S_{1}^{q,\text{virtual}}$, to the bare soft function are the same as in the gluon case up to Casimir scaling. Hence the rest of the analysis towards deriving the BFKL follows that of~\cite{Rothstein:2016bsq}, and we refer the reader there for further details. Let us mention a few key steps and then present the final evolution equation.
The rapidity divergence is multiplicatively renormalized with a $k_\perp$ convolution by a standard SCET soft function counterterm to cancel the $1/\eta$ divergence. Then the rapidity renormalization group follows from the $\nu$-independence of the bare soft function. The resulting RGE for $S^q(q_\perp, q^\prime_\perp)$ is precisely the leading log BFKL up to Casimir scaling:
\begin{align}
\nu \frac{d}{d\nu} S^q(q_\perp, q'_\perp,\nu)= \frac{2 C_F \, \alpha_s(\mu)}{\pi^2} \int d^2 k_\perp \left[  \frac{S^q(k_\perp, q'_\perp,\nu)}{(\vec k_\perp-\vec q_\perp)^2} -\frac{\vec{q}_\perp^{\,\,2} S^q(q_\perp, q'_\perp,\nu)}{2\vec k_\perp^2(\vec k_\perp-\vec q_\perp)^2} \right] \, .
\end{align}
Note that unlike the amplitude level Reggeization, the BFKL equation is IR finite due to the cancellation between the real and virtual emissions. 

Just as in \cite{Rothstein:2016bsq}, the rapidity RGE consistency,
\be
	0=\nu\frac{d}{d\nu} T^q_{(1,1)}  \implies 0 = \gamma_{S^q} + \gamma_{C_n^q} + \gamma_{C_\bn^q} = \gamma_{S^q} + 2\gamma_{C_n^q}\,,
\ee 
also implies a BFKL equation for the $n$-collinear function
\begin{align}
\nu \frac{d}{d\nu} C_n^q(q_\perp, p^-,\nu)= -\frac{ C_F \, \alpha_s(\mu)}{\pi^2} \int d^2 k_\perp \left[  \frac{C_n^q(k_\perp, p^-,\nu)}{(\vec k_\perp-\vec q_\perp)^2} -\frac{\vec{q}_\perp^{\,\,2} C_n^q(q_\perp, p^-,\nu)}{2\vec k_\perp^2(\vec k_\perp-\vec q_\perp)^2} \right] \,, 
\end{align}
and an analogous BFKL equation for $C^q_\bn$ with $(n,p^-,q_\perp)\leftrightarrow (\bn,p^{\prime +},q^\prime_\perp)$.

\section{Conclusions} \label{sec:conc}

In this paper we derived operators describing the exchange of Glauber quarks in the Regge limit, within the framework of the SCET.  
These Glauber quark operators describe certain soft and collinear gluon emissions to all orders in $\alpha_s$, and, for the case of a single soft gluon emission, reproduce the classic result of Fadin and Sherman~\cite{Fadin:1976nw,Fadin:1977jr}. From the rapidity renormalization of the Glauber quark operators, we derived the LL Reggeization of the quark and the LL BFKL equation for $q\bar q \to \gamma \gamma$. The rapidity renormalization gives rise to an interesting structure involving operator mixing between the $T$-product of two $\cO(\sqrt{\lambda})$ operators describing soft-collinear scattering, and an $\cO(\lambda)$ operator describing collinear-collinear scattering. We also showed that rapidity finite diagrams involving simultaneous Glauber quark and Glauber gluon exchanges quite simply reproduce known results in the $\bar 6$ and $15$ color channels, showing the consistency of our regulator. These results give a first view of the structure of the EFT for forward scattering in SCET at subleading power. 

There are a number of interesting directions for future study. In particular, it will be important to extend the study of Reggeization through renormalization group evolution to derive the two-loop Regge trajectory, both for the quark and the gluon. It is known that the two-loop quark Regge trajectory is related to the two-loop gluon Regge trajectory by Casimir scaling, $C_A \to C_F$~\cite{Bogdan:2002sr}, and it would be interesting to derive this property directly from the structure of Glauber operators, and to understand at what loop order it fails. Furthermore, now that the effective theory describes both quark and gluon Glauber exchanges, the structure of the higher logarithmic corrections for quantum numbers corresponding to compound Reggeon states can be studied using techniques in the effective theory. Finally, we have studied the subset of operators responsible for quark Reggeization at LL order, and it would be interesting to derive the complete set of power suppressed operators in the EFT for forward scattering, such as those describing subleading power corrections to the Regge trajectory of the gluon.

\vspace{0.2cm}
Note added: As this paper was being finalized, Ref.~\cite{Nefedov:2017qzc} appeared, which studies $\gamma \gamma \to q \bar q$ amplitudes at one-loop in the Regge limit by constructing the quark Reggeization terms in the effective action formalism of Lipatov~\cite{Lipatov:1995pn}. In the SCET language this corresponds to formulating an auxiliary field Lagrangian for the offshell Glauber quarks, while using the full QCD Lagrangian for other fields (without defining EFT fields for the $n$-collinear, soft and $\bar n$-collinear sectors).  Since having distinct fields for these sectors enables their factorization properties to be easily determined and studied, such as in our BFKL calculation, we believe there are certain advantages to our approach.  It would be interesting to make a more explicit comparison between these formalisms.

\begin{acknowledgments}
	
We thank Duff Neill, Ira Rothstein, and HuaXing Zhu for useful discussions. This work was supported in part by the Office of Nuclear Physics of the U.S. Department of Energy under the Grant No. DE-SCD011090, by the Office of High Energy Physics of the U.S. Department of Energy under Contract Numbers DE-AC02-05CH11231 and DE-SC0011632, and the LDRD Program of LBNL. I.S. was also supported in part by the Simons Foundation through the Investigator grant 327942.

\end{acknowledgments}

\appendix

\section{Expansions of Wilson Lines and SCET Conventions}\label{sec:conventions}

In this appendix we collect several expansions of Wilson lines and of the gauge invariant fields, which prove useful for deriving the Feynman rules used in the text. We use the following sign convention for the gauge covariant derivative
\begin{align}
G_{\mu \nu}^a=\partial_\mu A_\nu^a-\partial_\nu A_\mu^a+g f^{abc} A_{\mu}^b A_\nu^c\,,\qquad iD^\mu=i\partial^\mu+gA^\mu\,.
\end{align}
The collinear Wilson lines are defined by
\begin{align}
W_n =\left[  \sum\limits_{\text{perms}} \exp \left(  -\frac{g}{\bar \cP} \bar n \cdot A_n(x)  \right) \right]\,.
\end{align}
Expanded to two gluons with incoming momentum $k_1$ and $k_2$, we have
\begin{align}
W_n &=1-\frac{gT^a \bar n \cdot A_{nk}^a}{\bar n \cdot k}  +g^2 \left[   \frac{T^a T^b}{\bar n \cdot k_1 } +\frac{T^b T^a}{\bar n \cdot k_2}  \right]    \frac{\bar n \cdot A_{nk_1}^a   \bar n \cdot A_{nk_2}^b  }{2\bar n \cdot( k_1+ k_2)}+\cdots\,, \nn \\[3mm]
W^\dagger_n& =1+\frac{gT^a \bar n \cdot A_{nk}^a}{\bar n \cdot k}  +g^2 \left[   \frac{T^a T^b}{\bar n \cdot k_1 } +\frac{T^b T^a}{\bar n \cdot k_2}  \right]    \frac{\bar n \cdot A_{nk_1}^a   \bar n \cdot A_{nk_2}^b  }{2\bar n \cdot( k_1+ k_2)}+\cdots\,.
\end{align}
The collinear gluon field is defined as
\begin{align}
\cB^\mu_{n\perp}=\frac{1}{g}\left[   W_n^\dagger i D^\mu_{n\perp}W_n \right]\,.
\end{align}
Expanded to two gluons, both with incoming momentum, we find
\begin{align}
g\cB^\mu_{n\perp}&=g\left(   A^{\mu a}_{\perp k} T^a -k^\mu_\perp \frac{\bar n \cdot A^a_{nk} T^a}{\bar n \cdot k} \right)+g^2(T^a T^b-T^b T^a) \frac{\bar n \cdot A^a_{n k_1} A^{\mu b}_{\perp k_2} }{\bar n \cdot k_1}  \\
&+g^2 (k^\mu_{1\perp} +k^\mu_{2\perp}) \left(   \frac{T^a T^b}{\bar n \cdot k_1}+\frac{T^b T^a}{\bar n \cdot k_2 } \right)\frac{\bar n \cdot A_{n k_1}^a \bar n \cdot A^b_{n k_2}}{2\bar n \cdot( k_1+ k_2)} \,. \nn
\end{align}
In both cases, at least one of the gluons in the two gluon expansion is not transversely polarized.

For the soft Wilson lines, we have
\begin{align}
S_n &=1-\frac{gT^a n \cdot A_{sk}^a}{ n \cdot k}  +g^2 \left[   \frac{T^a T^b}{n \cdot k_1 } +\frac{T^b T^a}{n \cdot k_2}  \right]    \frac{ n \cdot A_{sk_1}^a   n \cdot A_{sk_2}^b  }{2 n \cdot( k_1+ k_2)}+\cdots\,, \nn \\[3mm]
S^\dagger_n& =1+\frac{gT^a n \cdot A_{sk}^a}{ n \cdot k}  +g^2 \left[   \frac{T^a T^b}{n \cdot k_1} +\frac{T^b T^a}{ n \cdot k_2}  \right]    \frac{ n \cdot A_{sk_1}^a    n \cdot A_{s k_2}^b  }{2 n \cdot( k_1+ k_2)}+\cdots\,.
\end{align}
and
\begin{align}
g\cB^\mu_{s(n)\perp}&=g\left(   A^{\mu a}_{\perp k} T^a -k^\mu_\perp \frac{ n \cdot A^a_{sk} T^a}{ n \cdot k} \right)+g^2(T^a T^b-T^b T^a) \frac{ n \cdot A^a_{sk_1} A^{\mu b}_{\perp k_2} }{ n \cdot k_1}  \\
&+g^2 (k^\mu_{1\perp} +k^\mu_{2\perp}) \left(   \frac{T^a T^b}{ n \cdot k_1 }+\frac{T^b T^a}{n \cdot k_2 } \right)\frac{ n \cdot A_{s k_1}^a  n \cdot A^b_{s k_2}}{2n \cdot( k_1+ k_2)} \,. \nn
\end{align}

When evaluating diagrams involving the soft Glauber operators, the following combination is also useful
\begin{align}
S_n^\dagger S_{\bar n}&=1+   gT^a\left(\frac{ n \cdot A_{sk}^a}{ n \cdot k}  - \frac{ \bar n \cdot A_{sk}^a}{\bar  n \cdot k}\right) -g^2 T^a T^b \frac{n\cdot A_s^a}{n\cdot k} \frac{\bar n \cdot A_s^b}{\bar n \cdot k}\nn \\
&+g^2 \left[   \frac{T^a T^b}{n \cdot k_1 } +\frac{T^b T^a}{n \cdot k_2}  \right]    \frac{ n \cdot A_{sk_1}^a   n \cdot A_{sk_2}^b  }{2n \cdot( k_1+ k_2)}+g^2 \left[   \frac{T^a T^b}{\bar n \cdot k_1} +\frac{T^b T^a}{\bar n \cdot k_2}  \right]    \frac{ \bar  n \cdot A_{s k_1}^a  \bar n \cdot A_{s k_2}^b  }{2\bar n \cdot( k_1+ k_2)}\,.
\end{align}

\bibliography{quark_reggeization_bib}{}
\bibliographystyle{JHEP}

\end{document}